\documentclass{pasj00}

\begin{document}
\SetRunningHead{Deguchi et al.}{Red Supergiants in Embedded Star Clusters}
\Received{2009/10/09}
\Accepted{2010/02/01; PASJ Ver. 1.2  Jan. 25, 2010}

\title{SiO and H$_2$O Maser Observations of Red Supergiants in Star Clusters Embedded in the Galactic Disk}

\author{Shuji \textsc{Deguchi}\altaffilmark{1}, Jun-ichi \textsc{Nakashima}\altaffilmark{2}} 
\author{Yong \textsc{Zhang}\altaffilmark{2}, Selina S. N. \textsc{Chong}\altaffilmark{2}, Kazutaka \textsc{Koike}\altaffilmark{1}}
\and 
\author{Sun \textsc{Kwok}\altaffilmark{2}} 
\altaffiltext{1}{Nobeyama Radio Observatory, National Astronomical Observatory,\\
       and Graduate University for Advanced Studies, \\
              Minamimaki, Minamisaku, Nagano 384-1305}    
\altaffiltext{2}{Department of Physics, University of Hong Kong, Pokfulam Rd, Hong Kong, China \vspace{0.5cm}}
\altaffiltext{}{\rm (PASJ 62, No. 2, the April 25 2010 issue  in press)}
\KeyWords{Galaxy: open clusters and associations: individual (Mercer et al.'s \#4, \#8, \#13, Stephenson \#2), --- masers ---
 stars: mass-loss --- stars: supergiants}

\maketitle

\begin{abstract}
We present the result of radio observations of red supergiants in the star cluster, Stephenson's \#2, and candidates for red supergiants
in the star clusters,  \citet{mer05}'s \#4, \#8, and \#13,  in the SiO and H$_2$O maser  lines. 
The Stephenson's \#2 cluster and nearby aggregation at the South-West contain 
more than 15 red supergiants. We detected one at the center of  Stephenson's \#2
and three in the south-west aggregation in the SiO maser line,  and three of these 4 were also detected in the H$_2$O maser line.
The average radial velocity of the 4 detected objects is 96 km s$^{-1}$,
 giving a kinematic distance of 5.5 kpc, which locates this cluster near the base of the Scutum-Crux spiral arm. 
We also detected 6 SiO emitting objects  associated with the other star clusters.    
In addition, mapping observations in the CO $J=1$--0 line toward these clusters
revealed that an appreciable amount of molecular gas still remains around Stephenson's \#2 cluster
in contrast to the prototypical red-supergiant cluster,  Bica et al.'s \#122.
It indicates that a time scale of gas expulsion differs considerably in individual clusters.
\end{abstract}

\section{Introduction} 
Massive young star clusters occasionally harbor a number of 
red supergiants with an initial mass of $\sim 10$--$ 15\ M_{\odot}$ \citep{sch70}, 
which are destined for a supernova explosion leaving behind a neutron star as a final product \citep{heg03}. 
Since  red supergiants exhibit high luminosity, they are easily identified at infrared wavelengths 
even if large interstellar extinction is assumed. 
They can be used as an indicator of age and mass of star clusters embedded in the Galactic disk \citep{lad03,fig06}.  
A red supergiant with $L= 10^5 \ L_{\odot}$ has  
a progenitor mass of about  $15 \ M_{\odot}$ \citep{sal99},
which is likely the upper limit of the initial masses of red supergiants in star clusters. 
   
\citet{fig06} measured the equivalent widths of the CO first overtone bands, and  identified 
14 M supergiants in a cluster in Scutum [\#122 of \citet{bic03a} =RSGC1]. 
They inferred the initial mass of this cluster to be $\sim 2$--$4 \times 10^4 M_{\odot}$ and the age between 7 and 12 Myr,
based on the color-luminosity diagram assuming the Salpeter's initial mass function.
\citet{nak06} detected SiO masers from 4 of these red supergiants. They found that
the velocity dispersion of the maser sources is equivalent to that of a cluster with a mass of $\sim 10^4 M_{\odot}$,
 and that the average radial velocity of maser sources indicates
a kinematic distance of 6.5 kpc to the cluster. This example demonstrated that SiO
masers are a useful tool for studying kinematics of star clusters as well as
mass-losing process  of massive stars at the final stage of stellar evolution.   

In this paper, we present the results of  SiO and H$_2$O maser searches for red supergiants 
in \#2 of \citet{ste90}, and candidates for  red supergiants in the other star clusters in the \citet{mer05}'s catalog.
The former cluster contains more than 10 red supergiants 
\citep{ste90}. A  later near-infrared (NIR) study found more supergiants and early type stars in the Stephenson \#2 cluster \citep{nak01}. 
More recently, \citet{dav08} spectroscopically identified
15 red supergiants in the Stephenson \#2 cluster and  nearby aggregation 5$'$ south-west from the Stephenson \#2 cluster. 
\citet{mer05} cataloged about 200 infrared star clusters lying in the Galactic disk 
using the Spitzer/GLIMPSE survey. 
Some of them include nearby bright middle-infrared (MIR) sources, which are likely red supergiants 
showing a typical color of SiO maser sources. However, since the color ranges of evolved stars
in the NIR and MIR bands are somewhat overlapped with those of young stellar objects,
it is not easy to discriminate young stars from evolved stars solely by the infrared colors. 
Detections of SiO masers are crucial for confirming the supergiant status 
of candidate stars in embedded clusters. We selected  three star clusters (\#4, \#8 and \#13)  as observing targets
from \citet{mer05}'s catalog: each selected clusters includes multiple candidates for red supergiants.
The cluster \#13 of \citet{mer05},  which is located about 6$^{\circ}$ north of Stephenson \#2, 
is  interesting because of an associated bright IR object (144 Jy at 12 $\mu$m). 

We have also mapped CO emission of Stephenson \#2 and several other clusters selected from the Mercer et al.'s catalog.
In addition, we mapped CO emission toward the prototypical red-supergiant star cluster in Scutum (RSGC1) for comparison. 
Massive star clusters are formed in giant molecular clouds \citep{lad03}, but 
the gas components of the clusters are eventually expelled and dissociated by radiation
of massive stars while the stars are at the main sequence of stellar evolution, 
and at the later phase by supernova explosion \citep{lad84}.  Dynamical evolution and dissolution of star clusters
depend strongly on gas expulsion process  (see for example, \cite{boi03}). Therefore,
it is interesting to know how much amount of CO gas does remain in the star clusters with red supergiants.

\section{Observations and results}
\begin{figure}
  \begin{center}
    \FigureFile(65mm,65mm){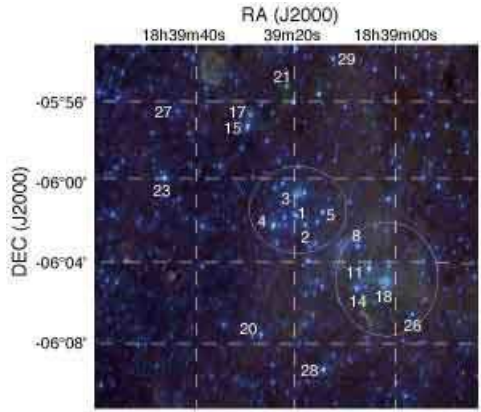}
  \end{center}
  \caption{a. Positions of the observed sources overlaid on the Spitzer/GLIMPSE image of Stephenson \#2. 
  The large two ellipses indicate approximate positions of the clusters, Stephenson \#2 (left) and \#2 SW (right).
}\label{fig:fig1a}
\end{figure}
\setcounter{figure}{0}
\begin{figure}
  \begin{center}
\FigureFile(60mm,60mm){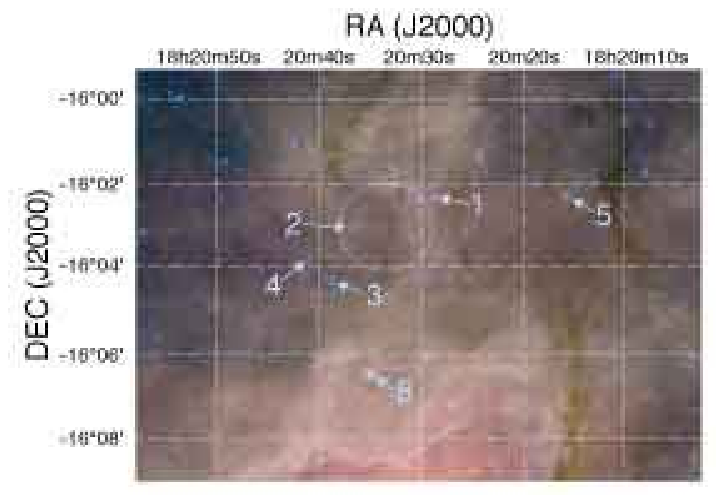}
  \end{center}
  \caption{b. Positions of the observed sources overlaid on the Spitzer/GLIMPSE image of Mercer et al.'s \#4 (shown by ellipse). 
}\label{fig:fig1b}
\end{figure}
\setcounter{figure}{0}
\begin{figure}
  \begin{center}
    \FigureFile(60mm,60mm){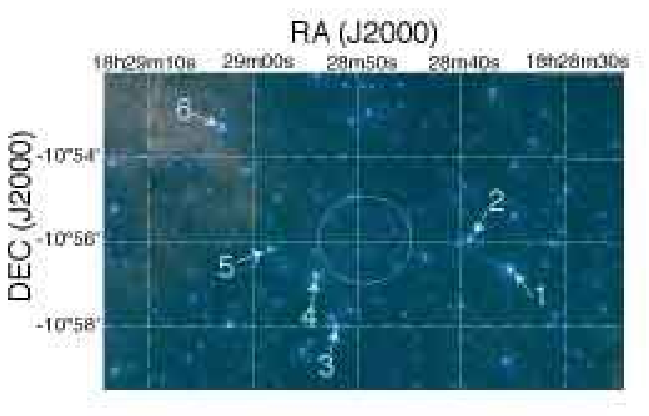}
  \end{center}
  \caption{c. Positions of the observed sources overlaid on the Spitzer/GLIMPSE image of Mercer et al.'s \#8  (shown by ellipse). 
}\label{fig:fig1c}
\end{figure}
%
\setcounter{figure}{0}
\begin{figure}
  \begin{center}
    \FigureFile(60mm,60mm){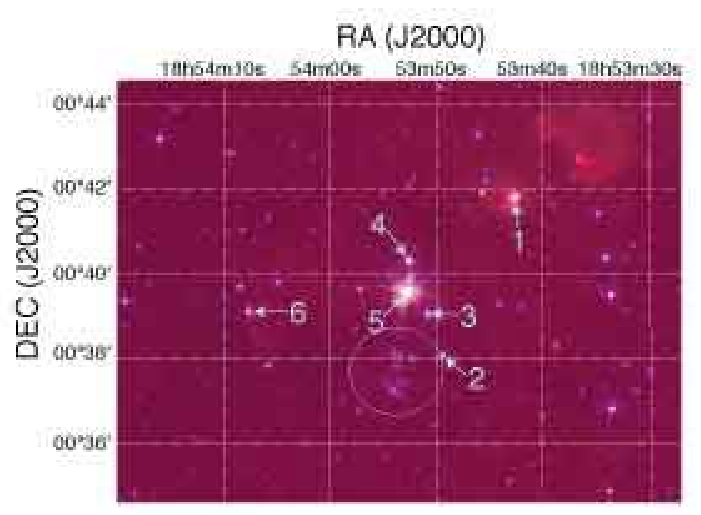}
  \end{center}
  \caption{d. Positions of the observed sources overlaid on the Spitzer/GLIMPSE of Mercer et al.'s \#13  (shown by ellipse). 
}\label{fig:fig1d}
\end{figure}

The observations were made with the 45m radio telescope at Nobeyama in 2006 April and 2008 April
in the H$_2$O $J_{KK}=6_{16}$--$5_{23}$ transition at 22.235 GHz,
the SiO $J=1$--0 $v=1$ and 2 transitions at 43.122 and 42.821 GHz, respectively,
and the CO $J=1$--0 transition at 115.271 GHz. 
Cooled HEMT receivers (H22 and H40)  were used for the 22 and 43 GHz observations
with acousto-opt spectrometer arrays with 40 and 250 MHz bandwidths (velocity resolution
of about 0.3 and 1.8 km s$^{-1}$ at 43 GHz, and 0.6 and 3.6 km s$^{-1}$ at 22 GHz, respectively).
The overall system temperature was about 140 K for 22 GHz H$_2$O observations, and
between 180 and 250 K for the 43 GHz SiO observations.
The half-power telescope beam width (HPBW) was about 90$''$ at 22 GHz and 40$''$ at 43 GHz. 
The conversion factor of the antenna temperature to the flux density was about 3.0 Jy K$^{-1}$ at 22 GHz,
and 2.9 Jy K$^{-1}$ at 43 GHz, respectively. 
All of the observations were made by the position-switching mode. Since the sources in the clusters are 
concentrated in a small region less than a $15'\times 15'$ area, we employed the position-switching sequence
like Off--On1--On2--On3, where the off position was taken 7$'$ west of the first on-source position. It 
saves the 50\% of integration time compared with a usual Off--On sequence \citep{deg04a}.
The $5\times 5$ beam focal-plane array receiver system (BEARS) was used 
for the 115 GHz CO observation (i.e., \cite{sun00}).
The system temperatures of the array receivers were between 400 and 600 K  (DSB).
The grid spacing of the array receiver was 41$''$  and the individual beam size was
 about 15$''$ (HPBW). The 1024 channel digital autocorrelator spectrometer \citep{sor00}
with a band width of 32 MHz was used for the 115 GHz observations, giving an effective velocity resolution
of 0.16 km s$^{-1}$ in a two-channel binding mode.
Further details of observations using the NRO 45-m telescope have been
described elsewhere\footnote{see http://www.nro.nao.ac.jp/~nro45mrt/obs/bears}, 
and are not repeated here. 

\begin{figure*}
  \begin{center}
    \FigureFile(150mm,180mm){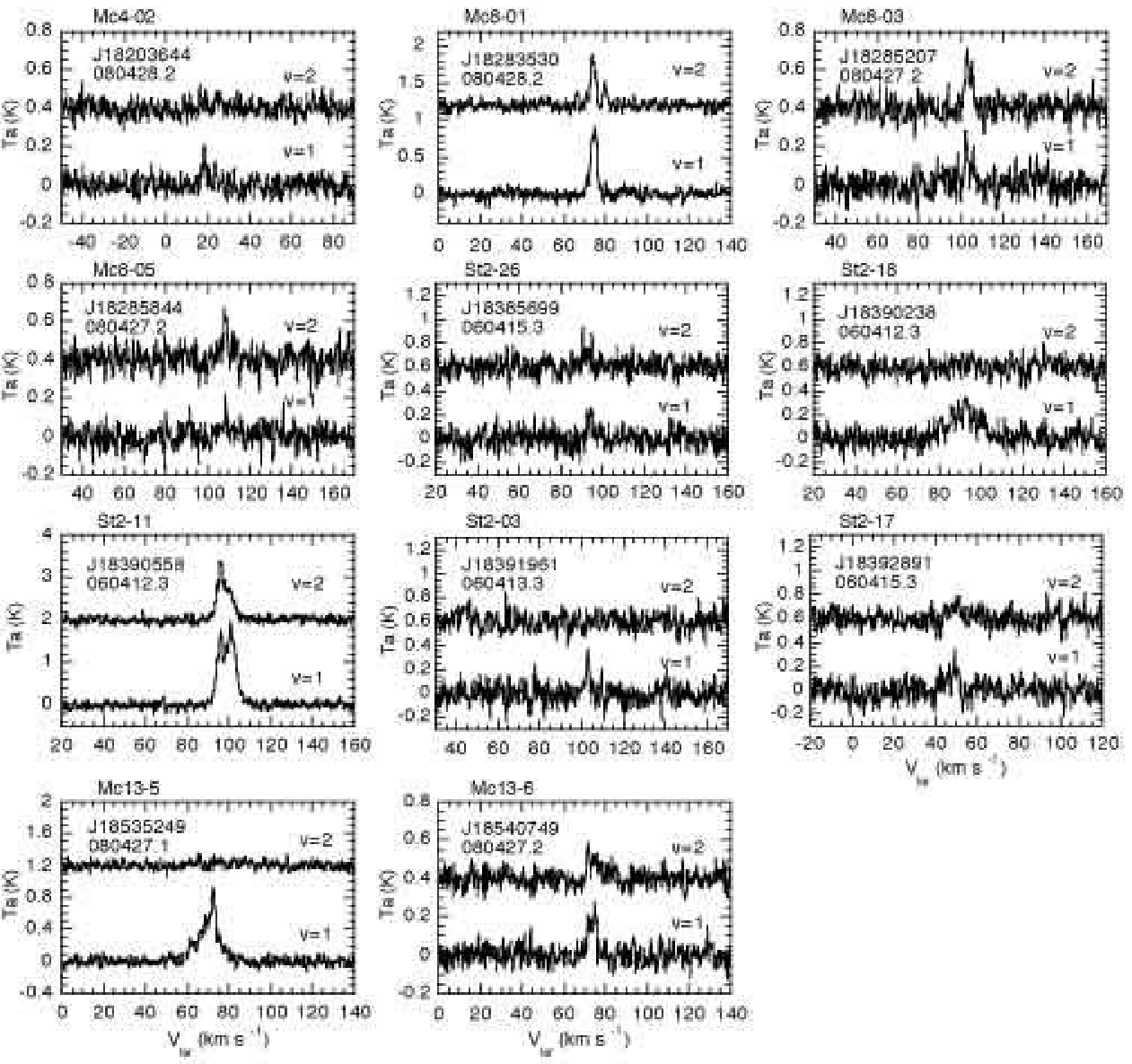}
  \end{center}
  \caption{a. SiO $J=1$-0 $v=1$ and 2 spectra. The star name,  
  abbreviation of 2MASS name (Jhhmmssss format), and the observed date (yymmdd.d format) are shown
   at the upper left of each panel.  
}\label{fig: spectra-1}
\end{figure*}
\setcounter{figure}{1}
\begin{figure*}
  \begin{center}
    \FigureFile(150mm,140mm){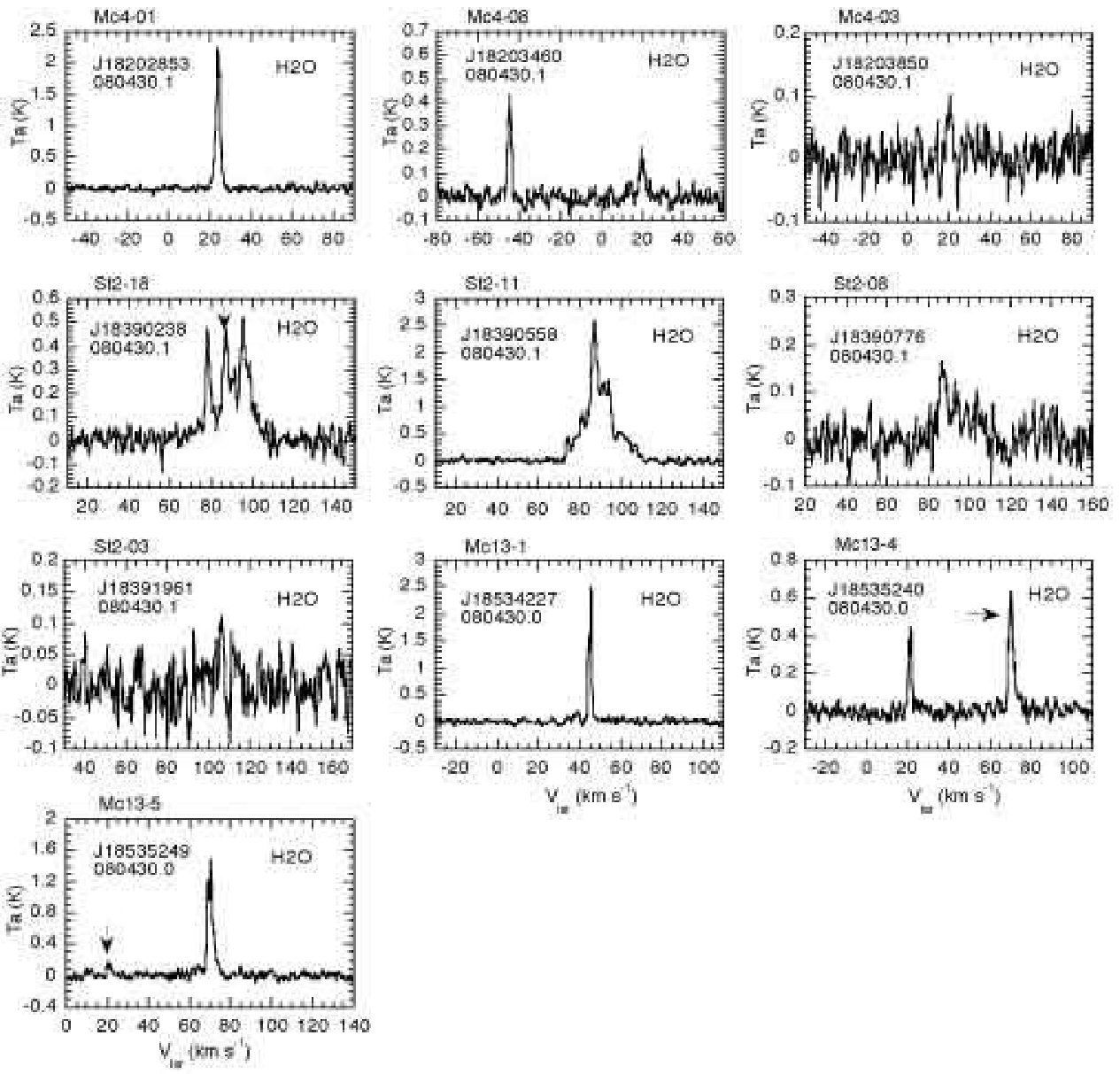}
  \end{center}
  \caption{b. H$_2$O $6_{16}$--$5_{23}$ spectra. The star name,  
  abbreviation of 2MASS name (Jhhmmssss format), and the observed date (yymmdd.d format) are shown
   at the upper left of each panel.   The arrows indicate the emission  
contaminated from the other object detected at the 
Gaussian tail of the beam. 
}\label{fig: spectra-2}
\end{figure*}

The data reductions were made using the NEWSTAR software package 
developed by Nobeyama radio observatory.
The BEARS data were first multiplied by the correction factors of antenna temperature 
for each beam displacement, and were integrated over time. Then 
the best fit quadratic curve was subtracted to determine the baseline of each spectrum.
Channel maps were created using the "MAP" task of NEWSTAR,
with a 2 km s$^{-1}$ separation and a 5 km s$^{-1}$ taper for channel smoothing.
The resulting CO channel maps were shown in Figures 5a--5f. 

Targets for SiO searches were chosen as follows.
We first selected  MIR objects brighter than 1 Jy in the MSX band C (12 $\mu$m) \citep{ega03}
toward the Stephenson \#2 and the Mercer et al.'s clusters. 
Since we noticed an aggregation of bright MIR objects 
5$'$ south-west of Stephenson \#2, we selected  objects from a relatively large area
 ($1000'' \times 1000 ''$) covering both Stephenson \#2 and the south-west aggregation,
 though some of them were possibly not a member of the cluster. 
We call the SW aggregation as Stephenson \#2 SW, hereafter.
Then, we checked to see if the selected MIR objects
have a NIR counterpart in a particular color range of the 2MASS colors \citep{cut03} 
(i.e.,  $K<9$, and $H-K>0.9$), which has been used as
criteria for target selections for our previous SiO maser surveys \citep{deg04b}.
Finally, we selected 18 infrared objects in the Stephenson \#2 and \#2 SW clusters.  
The positions of the selected objects were shown in Figure 1a. 

The star clusters with red supergiants other than the Stephenson \#2  were chosen
from the table of new star clusters discovered in the GLIMPSE survey data 
[Table 1 of \citet{mer05}]. We selected the MIR objects using the  criteria similar to the above.
Then, we eliminated several young stellar objects from this list by eye inspection
of 2MASS images. For example, those associated with large extended nebulosity 
(e.g., Mercer et al.'s \#40) were discarded.   
We selected 3 star clusters including 15 candidates for red supergiants.
In case of Mercer \#4, we added three red stars 
(J18202853$-$1602159, J18204070$-$1603412, 18203850$-$1604163),
which do not fill the criteria, but facilitate our observing method using a three-on-points sequence.
Table 1 summarizes the infrared properties of the observed sources.
Figure 1 shows the Spitzer/GLIMPSE images of the clusters and positions of the observed objects. 
The star clusters, Mercer et al.'s \#4 and \#8,  are a bit hard to recognize on the GLIMPSE images shown  in figures 1b
and 1c, respectively, because of the scales, but they can be seen clearly on enlarged 2MASS images
which are not shown here.  
For convenience, we assigned a short name to each star, such as
the abbreviated cluster name and the star number separated by a hyphen as  St2-01. 
Here, the Stephenson \#2 is presented by St2 (including St \#2 SW), and  Mercer et al.'s \#4,  \#8,  \#13 clusters
are presented by Mc4, Mc8, and Mc13, respectively. 
Table 4 summarizes the alternative naming of these objects.

Table 2 summarizes the observational results for SiO masers, and Table 3 for H$_2$O masers. 
Figures 2a and 2b show the SiO and H$_2$O maser spectra of detected sources, respectively. 
Figure 3 shows the K-magnitude versus NIR color and 12$\mu$m flux density versus MIR color diagrams, where
the detections and no detections are presented by filled and open circles, respectively.
Note that the beam size of the telescope  in the 22 GHz H$_2$O maser observation 
is 90$''$ (HPBW), approximately twice larger than the beam size in the 43 GHz SiO observation.
Since a mean separation between red supergiants in clusters is roughly 40--50$''$, 
we occasionally detected nearby H$_2$O maser sources which are contaminated from the Gaussian tail of the telescope beam. 
However, such contamination could be identified in the present observations,
because we observed multiple, close positions within one cluster.
The data gives us positional information on contaminating sources.

\subsection{Stephenson \#2}
Several bright objects are identified with IRAS sources: the brightest two sources are 
IRAS 18366$-$0603 (St2-03)  and 18363$-$0607 (St2-18) 
with $F_C =19.9$ and 56.9 Jy, respectively, and three others are
IRAS 18364$-$0605 (St2-08), 18368$-$0610 (St2-20), 18370$-$0602 (St2-23).
However, the other fainter sources have no IRAS counterpart. 
Though St2-26 is a faint object with $F_C=1.8$ Jy, 
 we detected SiO maser emission.  
The broad line profile of St2-18 suggests that this is a typical red supergiant
[for example, \citet{cer97}].

The radial velocity of St2-17 is shifted by more than 40 km s$^{-1}$ from the velocities of the other stars.
Therefore, we conclude that  St2-17 is not a member of Stephenson \#2.  We infer from this fact
that some stars at the northern part of Stephenson \#2 (i.e, St2-15, St2-21, St2-23, St2-27, and St2-29),
 which are located far from the cluster center, are not physically associated with Stephenson \#2. 
 
What is interesting is whether Stephenson \#2  SW  is
physically associated with the  Stephenson \#2. Note that 
Stephenson \#2 includes St2-01 up to St2-05, and  Stephenson \#2  SW
includes St2-08, St2-11, St2-14, St2-18 and St2-26.
The angular size of Stephenson \#2 suggested by \citet{bic03a} is $3.5' \times 3'$, 
and Stephenson \#2 SW is out of this area.
The average radial velocity and velocity dispersion for the SiO sources in Stephenson \#2 SW
are 94.5 km s$^{-1}$ and 3.5 km s$^{-1}$, respectively, which is shifted from the velocity of St2-3
in the Stephenson \#2 by 7.7 km s$^{-1}$. This difference is not large enough to rule out
the association of  Stephenson \#2 SW to Stephenson \#2. For convenience,
we presume that Stephenson \#2 SW is associated with Stephenson \#2.

\subsection{Mercer et al.'s \#4}
This cluster is located at the northern boundary of M17. SiO masers were detected only
in Mc4-02 (J18203644$-$1602589) at $V_{\rm lsr}=18$ km s$^{-1}$. Water masers
were detected at several locations. The radial velocity of Mc4-02 determined 
from the present SiO observation is close to that of the CO component of M17 
at $V_{\rm lsr}=20$--23 km s$^{-1}$ \citep{lad76}.
In fact, a strong CO emission is seen in the north-west parts of the channel maps 
in the range between $V_{\rm lsr}=20$ and 23 km s$^{-1}$.
The H$_2$O maser emission was detected toward three objects Mc4-01, Mc4-03 and Mc4-08.
Because of the large beam size at 22 GHz, H$_2$O emission of nearby objects 
could be detected at the tail of a Gaussian beam.
The velocity of H$_2$O  emission toward Mc4-03 is 20.5 km s$^{-1}$. This
is close to the velocity 19.4 km s$^{-1}$ of the secondary (redshifted) component of H$_2$O emission of Mc4-08. 
The angular separation between these two objects is 143$''$, 
indicating that the intensity should be reduced to 0.1\% level
if the 19.4 km s$^{-1}$ emission from Mc4-08 is detected at the Gaussian tail of the beam pattern at the position of Mc4-03. 
The observed ratio of the integrated intensity of Mc4-03 to that of Mc4-08 
is 29\%, which is  too large for a side-lobe detection, indicating that
the 20.5 km s$^{-1}$ emission of  Mc4-03  is not a contamination from Mc4-08
(though we cannot rule out the possibility that the emission comes from an unknown H$_2$O source 
located near the middle between Mc4-03 and Mc4-08).
Two water maser sources (possibly of a young star origin) have been detected in M17 north \citep{ces78,bra83}:
 G15.18$-$0.62 (+21 km s$^{-1}$), and G15.20$-$0.63 (+47 km s$^{-1}$).  They are located at
 20$''$ and 92$''$ away from Mc4-01 ($V_{\rm lsr}=24$ km s$^{-1}$).
Therefore, G15.18$-$0.62 could be the same H$_2$O maser source as Mc4-01, but G15.20$-$0.63 is 
different from Mc4-01.
The difference of 3 km s$^{-1}$ between peak velocities of Mc4-01 and G15.18$-$0.62 
could be explained by a time variation of a line profile. Thus,
we conclude that two of the three H$_2$O maser sources detected in the present
observation are new.

\subsection{Mercer et al.'s \#8}

 In Mc8-01,  SiO maser emission is detected at $V_{\rm lsr}=74$  km s$^{-1}$, which
 is considerably different from the average velocity 
of Mc8-03 and Mc8-04 (105.5 km s$^{-1}$). Therefore it is possible that the object Mc8-01
is a foreground source. Since  no water maser was detected toward this star cluster (Table 3),
dense molecular gas of a young star origin (as seen  toward  Mercer et al.'s \#4)
does not remain inside this star cluster.
In fact, CO channel maps shown in Figure 5d indicate that CO emission is quite weak toward this cluster,
except the 106 km s$^{-1}$ component, which is extended to the north-east of the star cluster
toward  Mc8-05.
A pulsar, PSR J1828$-$1057, exists $1'$ south-east of Mc8-0, and
three X-ray sources (XGPS-I J182844$-$105428 and XGPS-I J182851$-$105539,
GPSR5 20.722-0.075) exist in this direction.  The presence of these high-energy sources
seems consistent with the interpretation that CO gas was  wiped out by supernova explosions.

\subsection{Mercer et al.'s \#13}
Mc13-5 in this star cluster is a bright IR source ($F_C=145$ Jy), but
a search for SiO maser in this object was negative in a previous observation
\citep{hal90}. We detected the SiO masers for the first time in this object. 
Water maser was detected toward Mc13-1, Mc13-3, Mc13-4, and Mc13-5.
However, since the angular separation between Mc13-3 and Mc13-5 is only 41$''$, 
the H$_2$O emission at 70.2 km s$^{-1}$ detected at the position of Mc13-3 comes from Mc13-5.
Similarly, the angular separation between Mc13-4 and Mc13-5 is 46$''$, the emission at the position
of Mc13-4 is also a contamination from Mc13-05. The emission at 21.4 km s$^{-1}$
detected toward Mc13-4 is stronger than that detected at the position of Mc13-05.
Therefore, we conclude that the emission at 21.4 km s$^{-1}$ is associated with Mc13-04.  
A compact CO emission at $V_{lsr}=75$ km s$^{-1}$ was found at the position between
Mc13-5 and Mc13-6 (see Figure 5e), indicating that the molecular gas still remains in this cluster.

\subsection{Bica et al.'s \#122 =RSGC1}
In addition to the star clusters mentioned above, we observed 
CO emission toward \citet{bic03a}'s \#122 (=RSGC1 in \cite{dav07}),  which 
 involves 14 red supergiants \citep{fig06}. SiO masers have been detected
in four of them \citep{nak06}; the average radial velocity and velocity dispersion are
120 km s$^{-1}$ and 3 km s$^{-1}$, respectively. 
The lower right panel of Figure 4 shows the CO $J=1$--0 spectra of Bica et al.'s  \#122 
taken at 5 different positions aligning in the north--south direction
(centered toward the star F13; \cite{nak06}). CO emission was seen in the velocity range between
85 and 110 km s$^{-1}$. However,  no CO emission stronger than $T_a=0.2$ K was detected
at velocities above 110 km s$^{-1}$ toward this cluster in this work. This fact suggests
that almost no molecular gas remains in this cluster.  

\begin{figure}
  \begin{center}
    \FigureFile(90mm,70mm){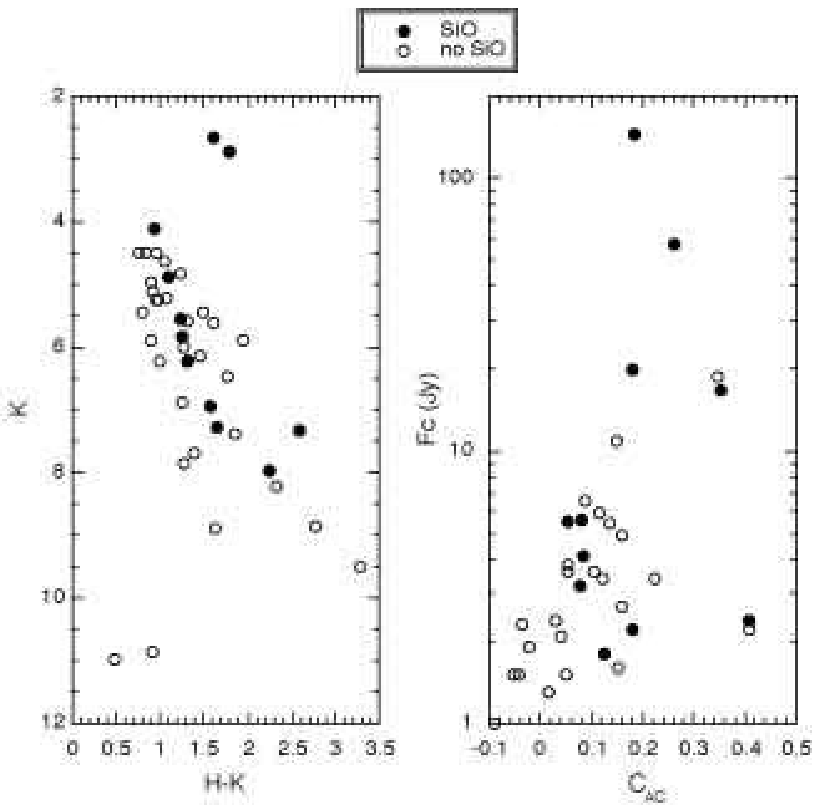}
  \end{center}
  \caption{$K$ magnitude versus $H-K$ color and 12 $\mu$m flux density versus $C_{AC}$ diagrams for the observed objects. 
Filled and unfilled circles indicate SiO maser detections and no detections, respectively.
}\label{fig: color-mag}
\end{figure}


\section{Discussion}
We observed 36 candidates for red supergiants in 4 star clusters in the SiO and H$_2$O  maser lines, 
and obtained accurate radial velocities.
The membership of stars to each cluster was confirmed by the velocity information obtained 
in the present observation,  since velocities of contaminated background stars 
are generally different from that of member stars. 
With the velocity information obtained by the present maser observations,
we can assess whether or not the CO emission is associated with the star cluster
in terms of radial velocities.
 
\subsection{Cluster kinematics of Stephenson \#2}
The age and distance of the cluster Stephenson \#2 were somewhat uncertain.
\citet{nak01} estimated the age and distance to be 50 Myr and 1.5 kpc, 
respectively, from a magnitude-color diagram. 
However,  \citet{ort02} suggested the age of 20 Myr and
a distance of 6 kpc for this cluster,  
based on the $VIJH$-band photometry and model calculations. 
The average radial velocity obtained in the present observation is 96.5 km s$^{-1}$,
 giving a kinematic distance of 5.5 kpc. This distance is consistent with 
the estimation given by \citet{ort02}. Here we assumed a flat rotation curve with a constant velocity of 220 km s$^{-1}$ 
and a distance to the Galactic center of 8 kpc. 

The standard deviation of radial velocities is 4.4 km s$^{-1}$ (for 4 stars). 
 To obtain this value, we assumed the uncertainty of 2 km s$^{-1}$ in the velocity measurement 
 of the SiO maser observation [see, Section 3.2 in \citet{nak06}]. 
 The velocity dispersion gives a virial mass of the cluster ($\equiv 6 {\bar r} {\bar v_r ^2 } /G $ ; \cite{eig04})
 of $1.3\times 10^5 M_{\odot}$ under the assumption that the radius of the cluster 
 (including both Stephenson \#2 and Stephenson \#2 SW) is 3.1$'$ ($\sim 5.5$ pc).
The obtained virial mass is  larger than the mass  of the Bica et al.'s \#122 (2--4$\times 10^4\ M_{\odot}$; 
\cite{fig06}). The virial mass only for Stephenson \#2 SW is  $3.7 \times 10^4\ M_{\odot}$ for a radius of
2$'$ (3.5 pc).  This value is close to the mass of  Bica et al.'s \#122, and
seems to be reasonable for a single massive cluster. 

\citet{dav07} obtained the K-band spectra of about 70 stars 
in Stephenson \#2 and  Stephenson \#2 SW with Kitt Peak National Observatory 4m telescope. 
They measured equivalent widths of the CO band-head features and obtained the radial velocities 
by cross-correlating the band head features with that of Arcturus. We compared the
radial velocities obtained in the present radio observation
with the velocities obtained by above infrared observations [see Figure 4 of \citet{dav07}].
The radio and infrared radial velocities of St2-18 [No 1 of \citet{dav07}] coincide within about 2 km s$^{-1}$,
 but other three stars do not show a good match in velocity, and the difference is roughly
8--10 km s$^{-1}$.  On average, the radial velocities obtained from the CO band head 
are systematically redshifted by about 10 km s$^{-1}$. 
It has been known that the radial velocities obtained from
infrared line profiles are systematically shifted from the true velocity of a star 
because of the scattering by dust grains 
in outflowing circumstellar envelopes \citep{rei76,van82}.  
Therefore,  if we take into account the uncertainty in the radial velocity obtained 
by the infrared observation, we can say that the velocities obtained 
in the present observation is consistent with those given by Davies et al. (2007).

\begin{figure*}
  \begin{center}
    \FigureFile(130mm,100mm){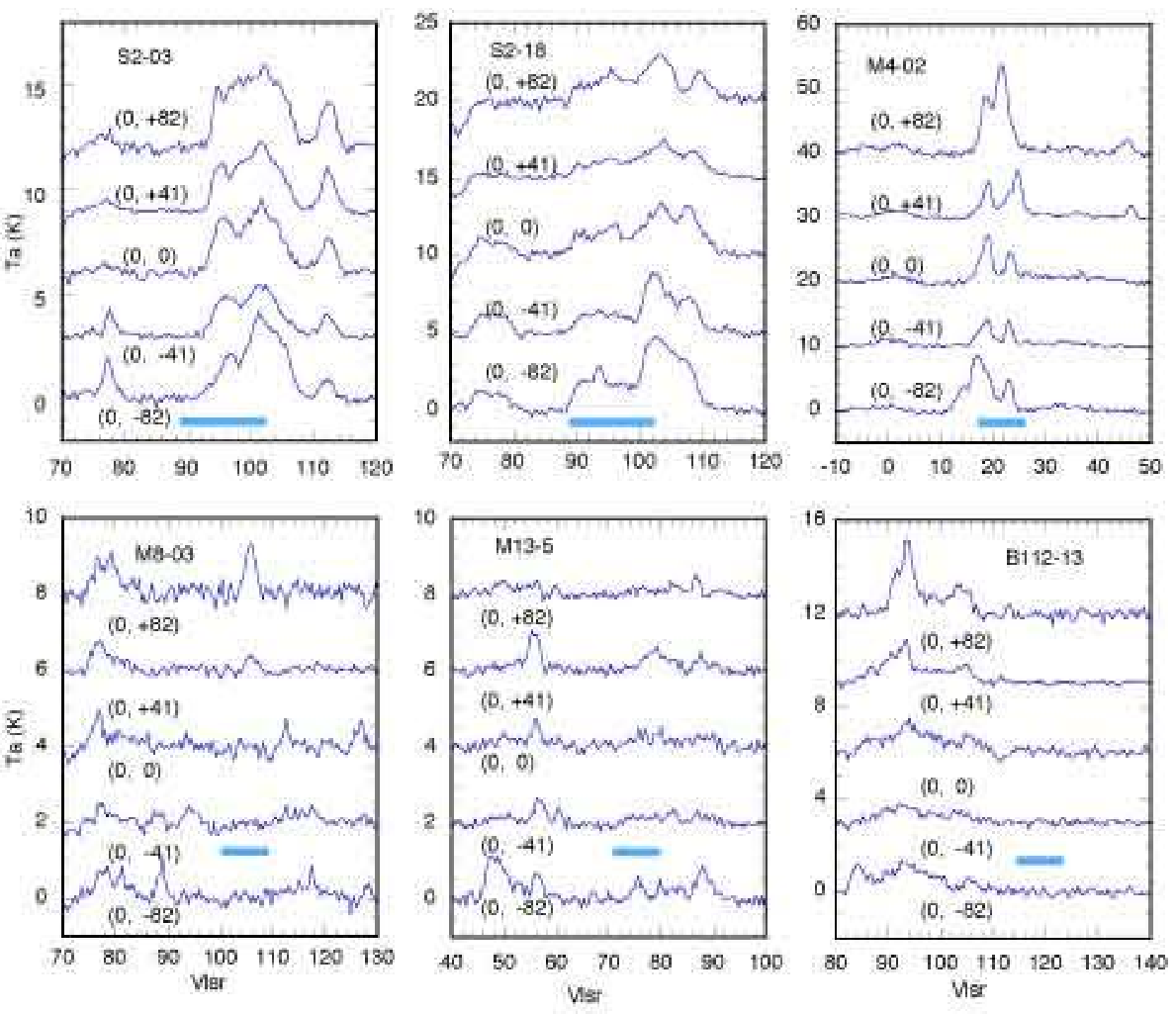}
  \end{center}
  \caption{CO $J=1$--0 spectra towards star clusters harboring observed targets. 
    The (0,0) positions correspond to the directions to St2-18, St2-03, Mc4-02, Mc8-03, Mc13-5.
	The (0,0) position in the last panel is toward 2MASS J18375890$-$0652321, 
				F13 of \cite{nak06} in Bica et al.'s \#122. 
    The offset positions were shown between the parentheses in arcsecond.
 The thick light blue bar in each panel indicates the possible velocity range 
 of star clusters.
}\label{fig: fig4}
\end{figure*}

\subsection{CO emission toward Stephenson \#2}
The CO channel maps in the velocity range  $V_{\rm lsr}=101$--105 km s$^{-1}$ (Figure 5a) 
show an emission enhancement near the center of the map 
(though the peak position does not coincide with the map center). 
The radial velocity of this cloud is approximately equal to the SiO radial velocity of St2-03
(102.7 km s$^{-1}$). Therefore it is likely that this cloud is associated with the star cluster
Stephenson \#2.  In contrast, the CO channel maps of Stephenson \#2 SW (Figure 5b)  
exhibit no strong enhancement of emission around the map center, 
but emission features are seen near the south edge of the map.
Additionally, weak emission peak  is noticeable at the north-east part of the channel maps 
 [(RA, Dec) $\sim (18^{\rm h}39^{\rm m}05^{\rm s}$, $-06^{\circ}04'30''$)]
in the $V_{\rm lsr}$ range between 93 and 103   km s$^{-1}$. 
The coordinates and velocity of this component
roughly coincide with those of St2-11. 
However, we attribute this CO emission to a fragment of the cloud toward this direction,
which is not related to the circumstellar envelope of the red supergiant St2-11, 
because the CO intensity is too strong for a red supergiant 5 kpc away.
 
\citet{jac06} mapped  the region  of $l=18^{\circ}$ -- $58^{\circ}$ and $|b|<1^{\circ}$ 
in the $^{13}$CO $J=1$--0 radio line with  the 22$''$ beam of on-the-fly sampling.
Their channel map at $V_{\rm lsr}=95$ -- 115  km s$^{-1}$
exhibits an emission peak  near Stephenson \#2 at $(l, b)\sim (26.15^{\circ}, -0.08^{\circ})$.
This position, possibly the densest part of the molecular cloud in this complex,
 is  a few arcminutes south of Stephenson \#2 and a few arcminutes east of Stephenson \#2 SW,
 but it is out of coverage of Figures 1a and 1b. 
If we assume that the kinematic distance is reliable,
 the Stephenson \#2 SW cluster ($V_{\rm lsr}\sim 94$  km s$^{-1}$ ) 
 is located at near side, and the thick CO clouds with the spectral peak showing an intensity peak 
 at $V_{\rm lsr}\sim 100$--105 km s$^{-1}$ is located at the far side of the Stephenson \#2 complex. 
Furthermore, Stephenson \#2  ($V_{\rm lsr}\sim 102$  km s$^{-1}$)  is close to the thick CO cloud. 
 This conjecture is consistent with the configuration of the
 Scutum-Crux arm, which has a tangency at distance of 7 kpc from the Sun  in the direction of $l\sim 35^{\circ}$.
 The  Scutum-Crux arm crosses the tip of the Bulge bar at distance $D\sim 6$ kpc in the direction of 
 $l\sim 30^{\circ}$, approaching toward the Sun with decreasing $l$  (see, e.g., \cite{ben08}). 
Therefore, the difference of the radial velocities between Stephenson \#2 
and Stephenson \#2 SW can be understandable in terms of the structure 
and kinematics of the Scutum-Crux arm.

\setcounter{figure}{4}
\begin{figure*}
  \begin{center}
    \FigureFile(100mm,150mm){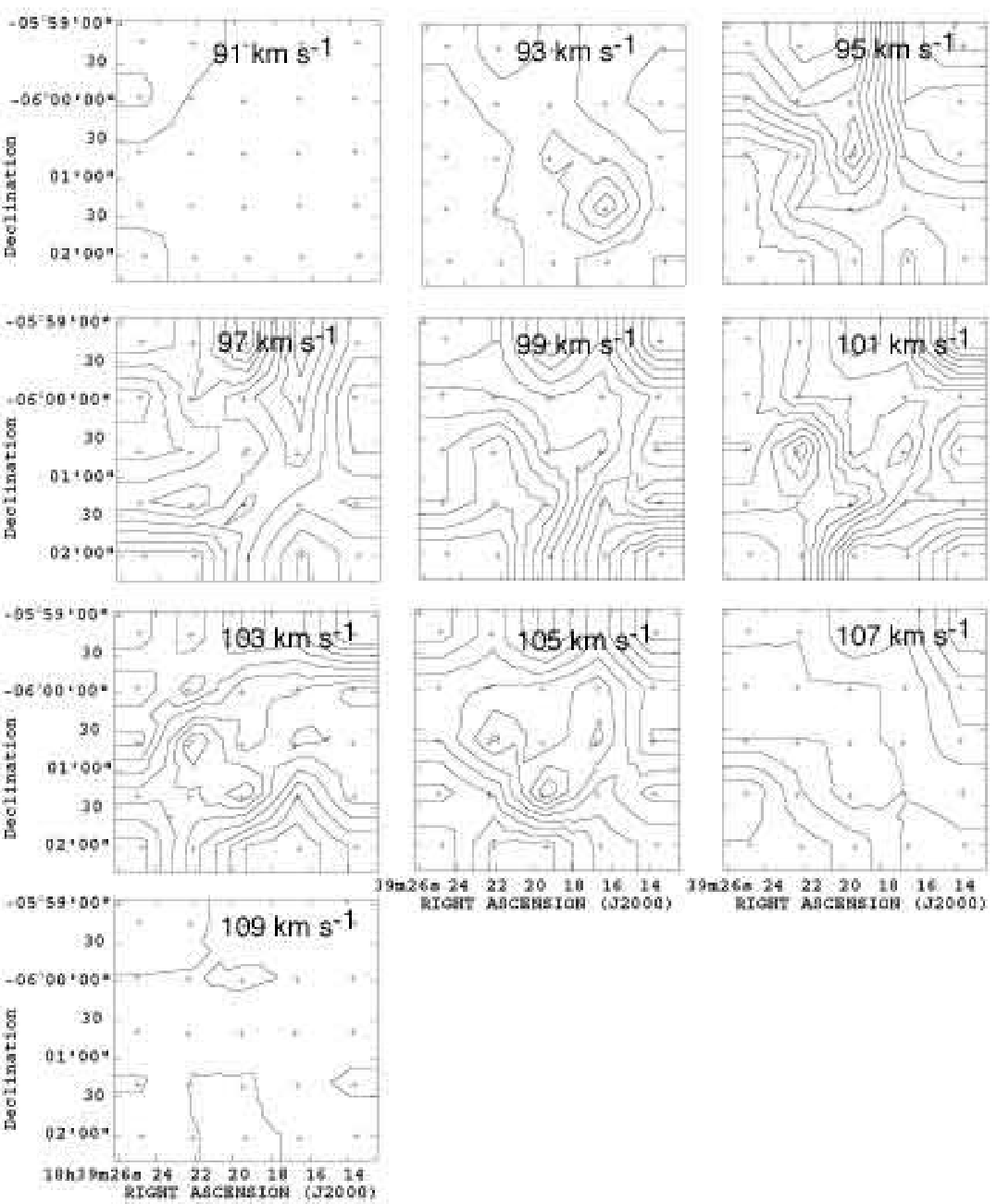}
  \end{center}
\caption{a. CO $J=1$--0 channel maps toward Stephenson \#2. Cross marks are observed positions by the array receiver.
The contour levels are drawn by every 0.3 K from the lowest one (0.3 K)
}\label{fig: fig5a}
\end{figure*}
\setcounter{figure}{4}
\begin{figure*}
  \begin{center}
    \FigureFile(100mm,150mm){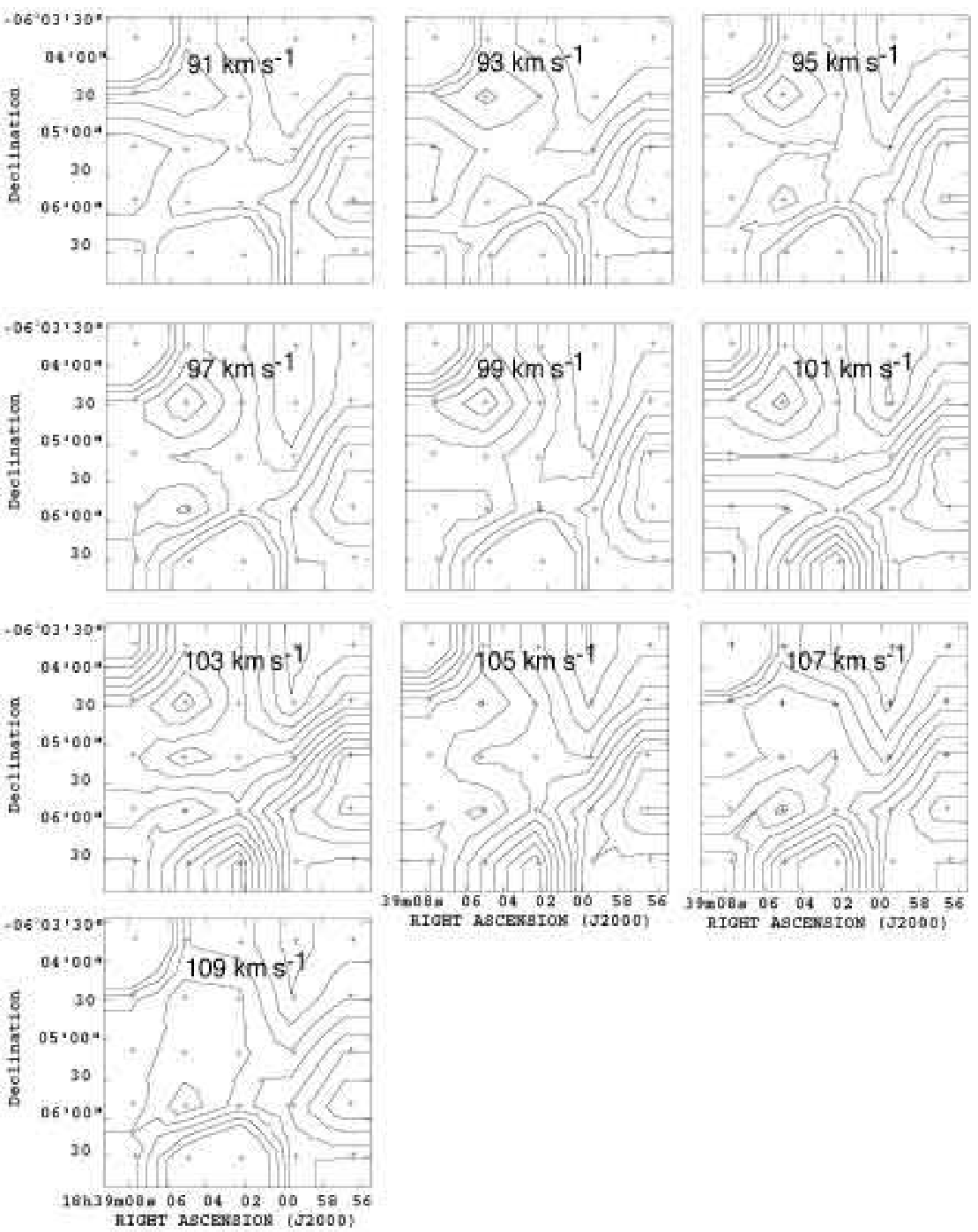}
  \end{center}
  \caption{b. CO $J=1$--0 channel map toward Stephenson \#2 SW. Cross marks are observed positions by the array receiver.
The contour levels are drawn by every 0.5 K from the lowest one (0.5 K)
}\label{fig: fig5b}
\end{figure*}

\subsection{CO toward the Bica et al.'s \#122 and comparison with Stephenson \#2}
The red-supergiant star cluster, Bica et al.'s \#122 (RSGC1), is located very close 
to Stephenson \#2 with a separation of about 1.0$^{\circ}$. 
No CO emission is seen toward Bica et al.'s \#122  at the velocities above 110 km s$^{-1}$ in
lower right panel of Figure 4 and in Figure 5f.  This fact suggests 
that almost no molecular gas is left behind Bica et al.'s \#122.
This is a sharp contrast with the case of Stephenson \#2 (RSGC2). 
The ages of  red supergiants of Bica et al.'s \#122 and Stephenson \#2 were estimated to be 8 and 17 Myr,
respectively, from isochrone fittings using the "fast-rotating" Geneva models. 
\citet{dav07} found that ages of red supergiants in Stephenson \#2 show
a relatively large dispersion (several Myr), and therefore they concluded that
the red supergiants in Stephenson \#2 and \#2 SW are not coeval.
For the case of Stephenson \#2, the radial velocity obtained by the present SiO maser observation 
is slightly different from that obtained by the NIR observation. 
This fact suggests that the distance to Stephenson \#2 and \#2 SW clusters is closer than the previous expectation.
 It is safely concluded 
that Stephenson \#2 and \#2 SW is located in front of the base of the Scutum-Crux arm
(e.g., \cite{rus03}), and  that Bica et al.'s \#122 is much farther than Stephenson \#2;
i.e.,  Bica et al.'s \#122 is located near the tangent point of this direction, 
 possibly inside the circle of the Galactic bulge tip.
The aged cluster Stephenson \#2 appears to hold more CO gases 
than the younger cluster Bica et al.'s \#122 does.
This fact seems to indicate that the gas expulsion
from a star cluster is not a simple function of time, and it depends on the
environment of the cluster, possibly on a mass and sphericity of the original molecular cloud,
a distribution of birth dates of massive stars, and a positioning of sequence of cloud formation
in a spiral arm. 

Note that the radial velocity of Stephenson \#2 ($\sim 96$ km s$^{-1}$) falls 
on the "Molecular Ring" feature in the longitude-velocity map 
of the Galactic disk CO emission [see, \citet{dam01}], 
but the velocity of Bica et al.'s \#122 (120 km s$^{-1}$) exceeds the molecular ring velocity at $l=26^{\circ}$.
This fact suggests that  Bica et al.'s \#122 is formed  under strong influence of the Galactic bulge
unlike majority of usual star clusters formed in the Galactic spiral arms.

\setcounter{figure}{4}
\begin{figure*}
  \begin{center}
    \FigureFile(80mm,100mm){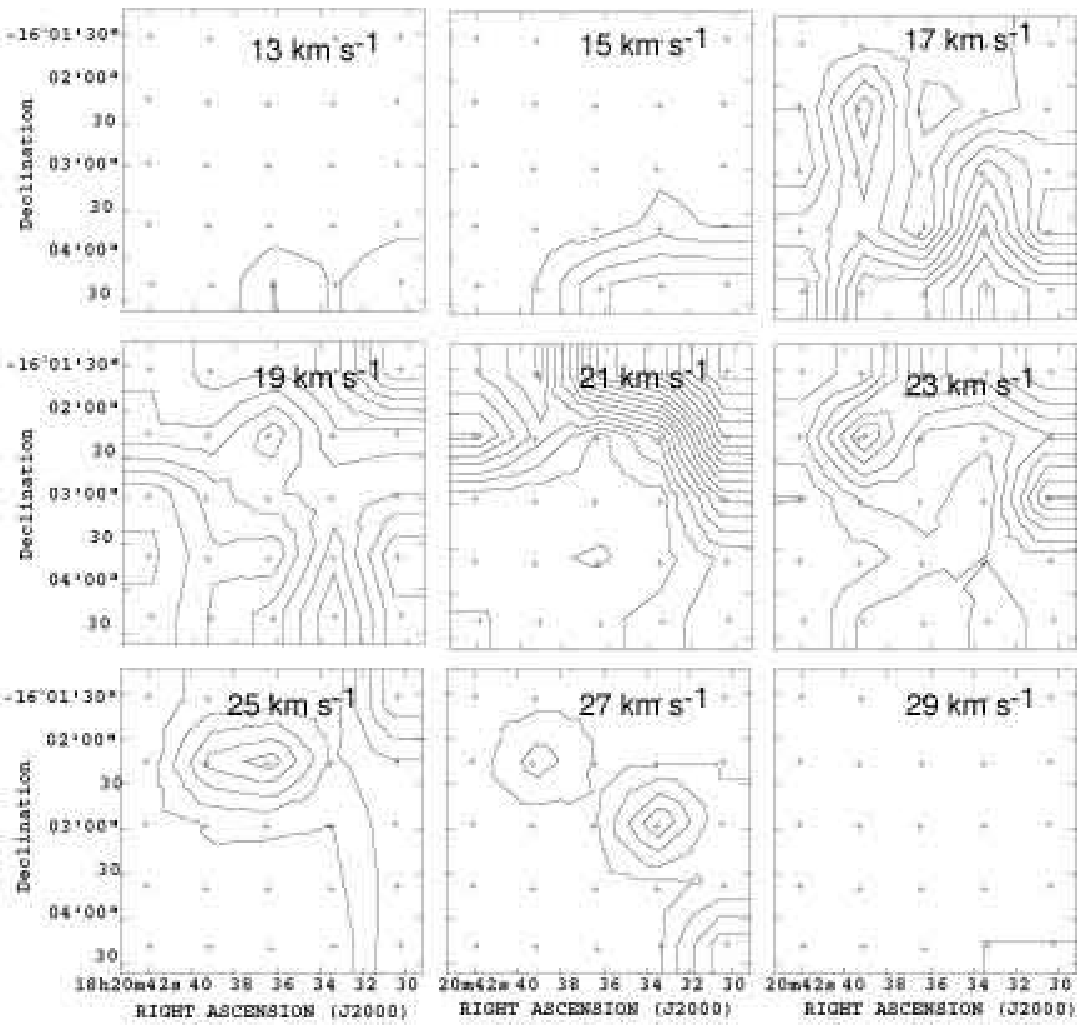}
  \end{center}
  \caption{c. CO $J=1$--0 channel map toward Mercer et al.'s \#4. Cross marks are observed positions by the array receiver.
The contour levels are drawn by every 1.0 K from the lowest one (1.0 K)
}\label{fig: fig5c}
\end{figure*}
\setcounter{figure}{4}
\begin{figure*}
  \begin{center}
    \FigureFile(80mm,100mm){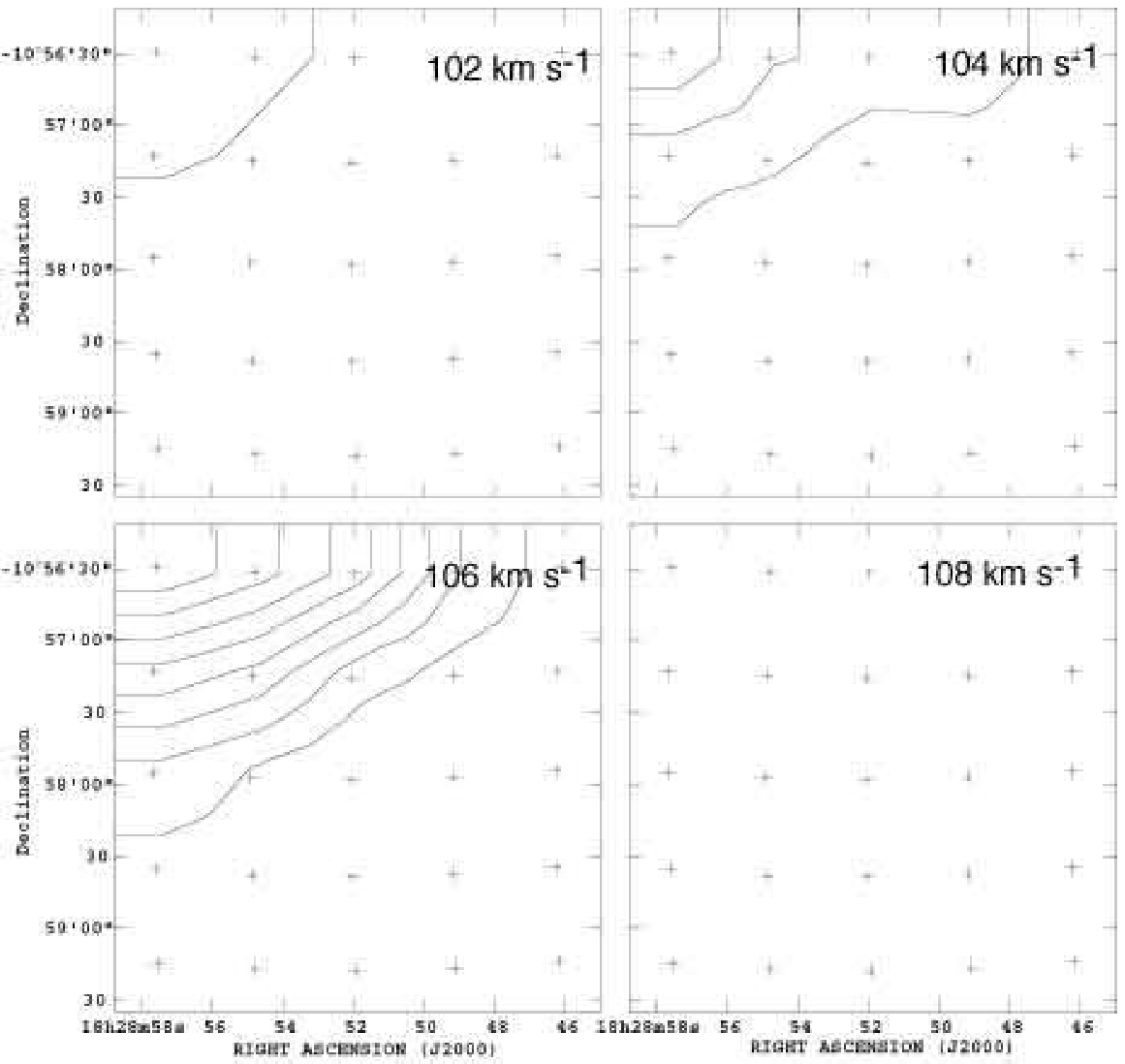}
  \end{center}
  \caption{d. CO $J=1$--0 channel map toward Mercer et al.'s \#8. Cross marks are observed positions by the array receiver.
The contour levels are drawn by every 0.2 K from the lowest one (0.2 K)
}\label{fig: fig5d}
\end{figure*}
\setcounter{figure}{4}
\begin{figure*}
  \begin{center}
    \FigureFile(80mm,100mm){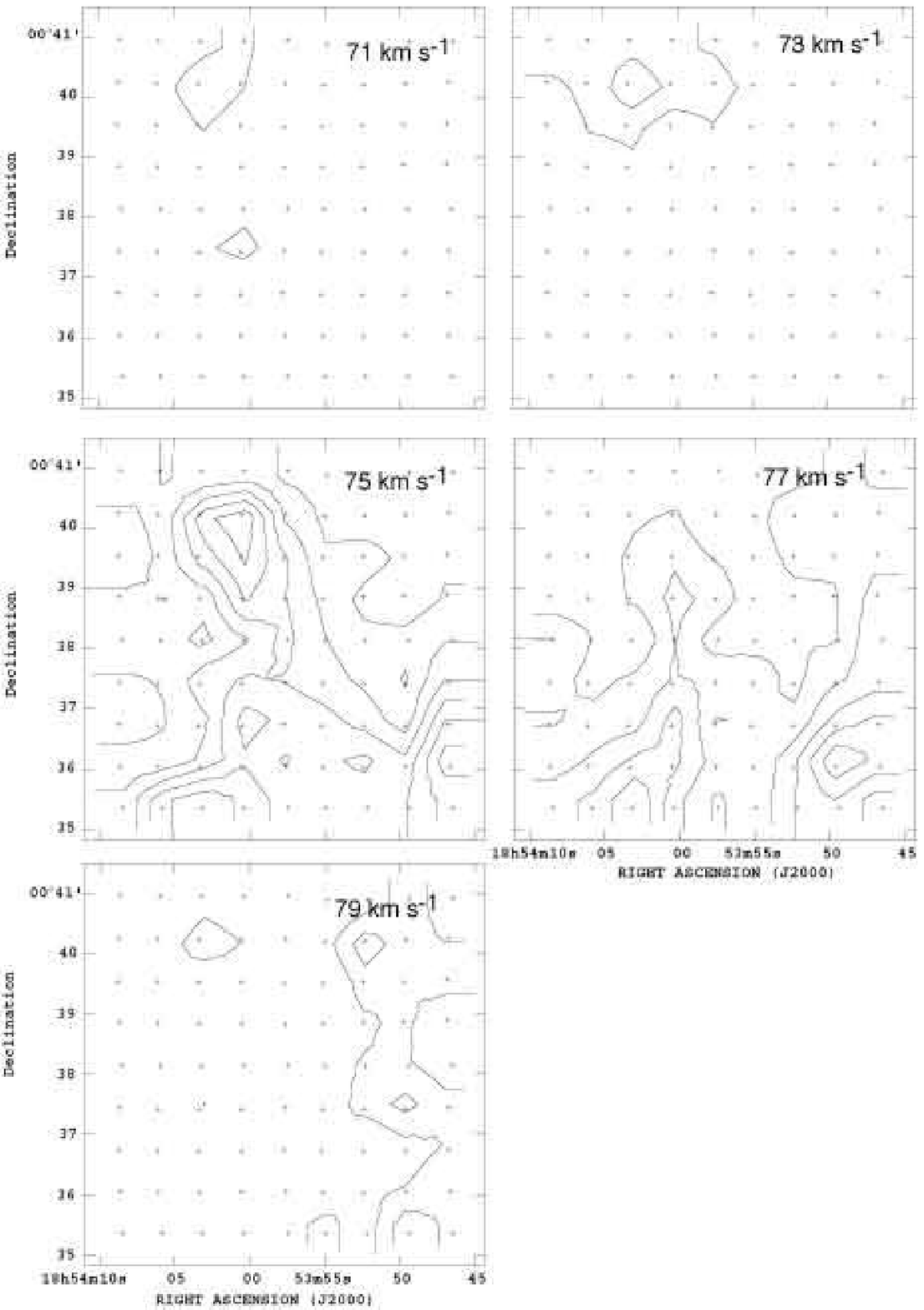}
  \end{center}
  \caption{e. CO $J=1$--0 channel map toward Mercer et al.'s \#13. Cross marks are observed positions by the array receiver.
The contour levels are drawn by every 0.2 K from the lowest one (0.2 K). 
  from the left to the right.
}\label{fig: fig5e}
\end{figure*}
\setcounter{figure}{4}
\begin{figure*}
  \begin{center}
    \FigureFile(80mm,100mm){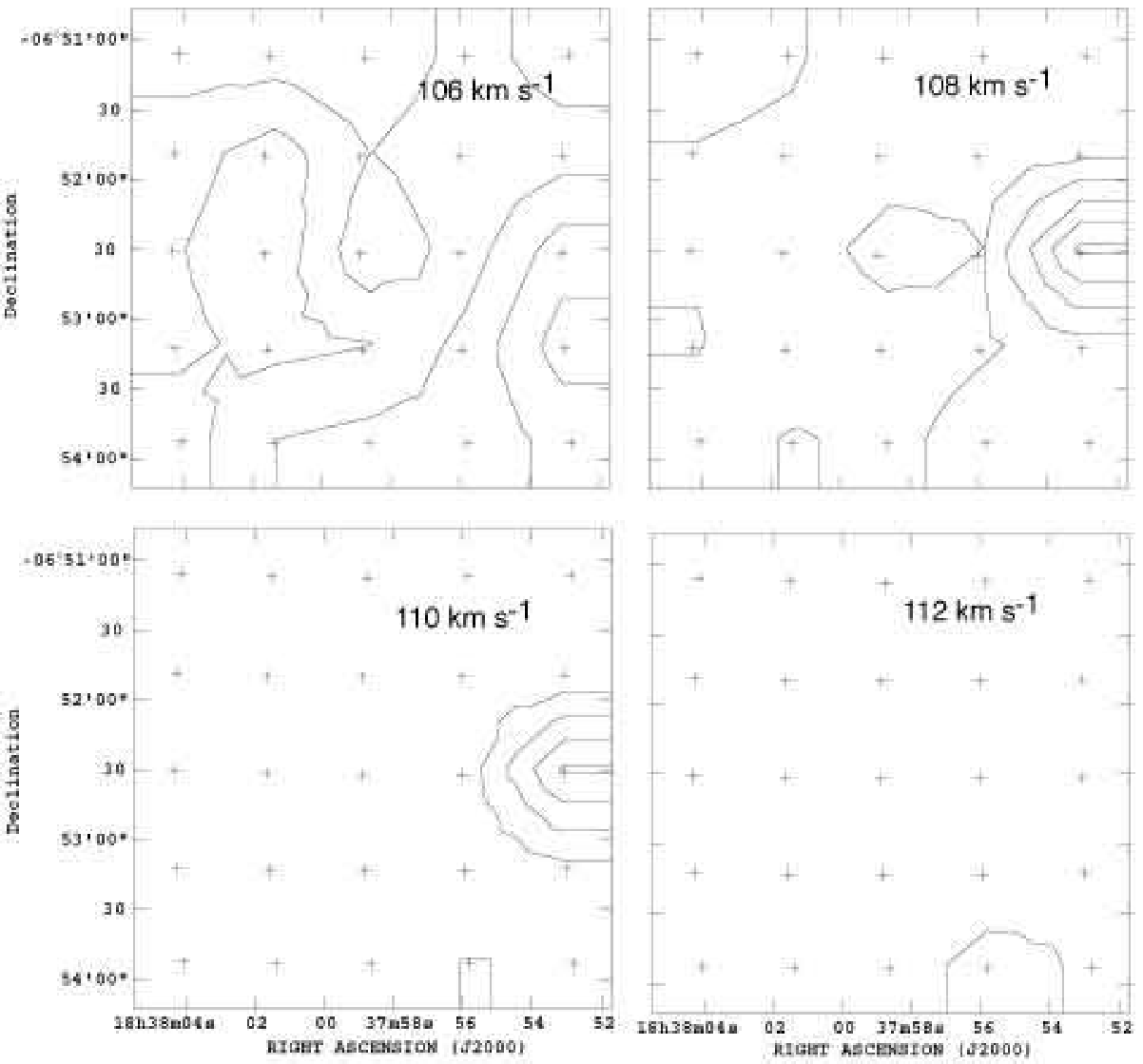}
  \end{center}
  \caption{f. CO $J=1$--0 channel map toward Bica et al.'s \#112 (=RSGC1). Cross marks are observed positions by the array receiver.
The contour levels are drawn by every 0.2 K from the lowest one (0.2 K).
No CO emission stronger than $T_a=0.2$ K was found in channels of $V_{\rm lsr}>112$ km s$^{-1}$, . 
}\label{fig: fig5f}
\end{figure*}

\section{Summary}
We detected SiO maser emission ($J=1$--0 $v=1$ and 2) toward
11 of 36 observed targets including red supergiants 
and candidates for red supergiants in star clusters embedded in the Galactic disk.
We also detected H$_2$O maser emission ($J_{KK}$=6$_{16}$--5$_{23}$) toward 10 of these 36 targets, 
in which  both SiO and  H$_2$O  lines were detected in 4. 
Accurate radial velocities of the detected objects were obtained from velocities of SiO maser lines.
The estimated kinematic distance  
indicates that the distance of Stephenson \#2
  is significantly smaller than that of  Bica et al.'s \#122. 
The radial velocity obtained by the present observation was used to check the association of  CO gas
to these clusters.
We found that a plenty of molecular gas still remains in  Stephenson \#2, whereas
almost no molecular gas remains in  Bica et al.'s \#122.  
We conclude that the SiO and H$_2$O masers are useful tools for investigating the
massive stars in star clusters and  kinematics of star clusters,
as well as processes of mass loss of red supergiants.  
At the end,  these studies will  clarify  detailed  processes of disruption mechanisms of
massive star clusters in the Galaxy.

\

This work is supported in part by Grant-in-Aid for Scientific Research 
from Japan Society for Promotion of Sciences (20540234),
and by a grant awarded to JN, YZ, and SK from the
Research Grants Council of Hong Kong (project code: HKU 703308P) and the
Seed Funding Program for Basic Research of the University of Hong Kong
[project code: 200802159006 (JN) and 200909159007 (YZ)], and by a grant from the Research 359
Grants Council of the Hong Kong Special Administrative 360
Region, China (Project No.HKU 7020/08P).
This research made use of the SIMBAD and VizieR databases operated at CDS, 
Strasbourg, France, and as well as use of data products from 
Two Micron All Sky Survey, which is a joint
project of the University of Massachusetts and Infrared Processing 
and Analysis Center/California Institute of Technology, 
funded by the National Aeronautics and Space Administration and
National Science foundation, and from the Midcourse Space 
Experiment at NASA/ IPAC Infrared Science Archive, which is operated by the 
Jet Propulsion Laboratory, California Institute of Technology, 
under contract with the National Aeronautics and Space Administration.   

\section*{Appendix. Individually interesting objects}

\begin{itemize}
\item Mc4-02  (=M17WF 26): 
This star is located near the north boundary of M17. \citet{and02} studied the infrared objects around M17
and recorded this star as \#26 in their Table 2 (also see their Figure 3). 
We detected SiO maser emission at $V_{\rm lsr}=18$ km s$^{-1}$.
Corrected K magnitude of this star is $K_c$=5.45. If this object is at the same distance 
as that of M17 (2.4) kpc \citep{lad76}, the luminosity is estimated to be $\sim 10^3 \ L_{\odot}$, 
which is too small as a supergiant. The $K_s$ magnitude of this star was measured in the past 
as $7.98\pm 0.02$ (2MASS), $8.15\pm 0.05$ (DENIS), and $8.2\pm 0.2$ \citep{and02}.
This star is possibly a distant bulge population, which has no direct relation to Mercer \#4. 

\item Mc8-01 (=IRAS 18258$-$1058):
The SiO masers were detected in this star by \citet{izu99} in their  SiO maser survey of Galactic bulge stars. 
 \citet{deg98} suggested a questionable NIR counterpart  located 8$''$ south of the IRAS position.
However, the 2MASS star, J18283530$-$105636, which is identified in the present work, 
is much brighter, and is spatially closer to the MSX counterpart (G020.6034+00.0234).
The SiO radial velocity ($\sim 74$ km s$^{-1}$) 
suggests that this object is not the member
of  Mercer et al.'s \#8, and is possibly associated with the Galactic bulge. 
 
\item Mc13-1 (=IRAS 18511+0038): 
This object is an ultra-compact HII region \citep{giv07}. 
In fact, CS $J=7$--6 emission was detected toward this source at $V_{\rm lsr}=46$ km s$^{-1}$ \citep{plu92}, 
though a search for the CS $J=1$-0 and NH$_3$ (1,1) lines was negative \citep{ang96}. 
Water maser emission has been detected by \citet{bra83} at $V_{\rm lsr}=38$ km s$^{-1}$,
which is close to the peak velocity of 45.4 km s$^{-1}$ obtained in this work. 
Since the large difference is found in the radial velocities,
this object is not associated with Mercer et al.'s \#13 which exhibits a radial velocity of
$\sim 72$ km s$^{-1}$. No SiO maser was detected in this star.

\item St2-03 (=IRAS 18366$-$0603) exhibits a relatively flat IRAS LRS spectrum 
with a small hump around 10 $\mu$m \citep{kwo97}.
SiO and H$_2$O maser emissions were detected at $V_{\rm lsr}=103$ and 106 km s$^{-1}$, 
respectively, in the present observation.
A search for the OH 1612 MHz maser was negative in a previous observation [IRC $-$10447;   \cite{wil72}].

\item St2-18 (=J18390238$-$0605106): 
This is an enigmatic object showing a K magnitude of 1.03 (corrected for interstellar extinction).
 IRAS LRS spectrum (IRAS 18363$-$0607) exhibits 
strong 10 and 18 $\mu$m silicate emission \citep{vol89}.
 Note that the typical $K$ magnitude of SiO sources (AGB stars) 
in the Galactic bulge  is about 5.5 \citep{fuj06}. It indicates the luminosity of St2-18
 is 30 times the luminosity of average bulge SiO sources at the K bands.
An estimation of luminosity using the IRAS 12 and 25 $\mu$m flux densities
\citep{jia96} gives the luminosity of $9 \times 10^4\ L_{\odot}$
at the distance of 5.5 kpc.
\end{itemize}


\clearpage
\renewcommand{\arraystretch}{0.9}
\begin{longtable}{lcrrrrrrrrr}
  \caption{Properties of observed objects in Stephenson and embedded clusters.} 
\hline  \hline
2MASS name       & No$^{*}$ & $K$ & $J-H$ & $H-K$ &  MSX name & sep & $F_c$  &  $C_{AC}$ & $C_{CE}$ & $V_{\rm lsr}^{SiO}$ \\ 
                 &        &     &       &       &           & ($''$) & (Jy)  &           &          & {\tiny (km s$^{-1}$)}  \\ 
\hline
\endfirsthead
\hline \hline
2MASS name       & No$^{*}$ & $K$ & $J-H$ & $H-K$ &  MSX name & sep & $F_c$  &  $C_{AC}$ & $C_{CE}$ & $V_{\rm lsr}^{SiO}$ \\ 
                 &        &     &       &       &           & ($''$) & (Jy) &           &          & {\tiny (km s$^{-1}$)}  \\ 
\hline
\endhead
\hline
\endfoot
\hline
\multicolumn{11}{l}{$^{*}$ the star number in figure 1.}\\
\endlastfoot
18201544$-$1602142 & Mc4-05 & 8.23 & 4.19 & 2.32 & G015.1542$-$00.5707 & 0.8 & 2.2 & 0.408 & 0.034 & --- \\ 
18202853$-$1602159 & Mc4-01 & 7.38 & 3.50 & 1.86 & ---                 & ---  & --- & --- & --- & --- \\
18203460$-$1606282 & Mc4-08 & 8.89 & 2.35 & 1.63 & G015.1288$-$00.6717 & 3.0 & 10.9 & 0.150 & 0.917 & --- \\ 
18203644$-$1602589 & Mc4-02 & 7.98 & 4.04 & 2.26 & G015.1831$-$00.6482 & 8.6 & 2.4 & 0.407 & -0.247 & 18.3 \\ 
18203850$-$1604163 & Mc4-03 & 7.85 & 2.58 & 1.27 & --- & ---  & --- & --- & --- & --- \\ 
18204070$-$1603412 & Mc4-04 & 7.70 & 2.86 & 1.39 & --- & ---  & --- & --- & --- & --- \\ 
18283530$-$1056364 & Mc8-01 & 4.88 & 1.92 & 1.10 & G020.6034$+$00.0234 & 0.4 & 5.6 & 0.081 & -0.184 & 74.1 \\ 
18283912$-$1055526 & Mc8-02 & 4.97 & 1.89 & 0.89 & G020.6214$+$00.0151 & 1.1 & 1.0 & -0.091 & -0.154 & --- \\ 
18285207$-$1057576 & Mc8-03 & 6.23 & 2.23 & 1.32 & G020.6156$-$00.0476 & 1.7 & 4.1 & 0.085 & -0.105 & 102.4 \\ 
18285402$-$1056453 & Mc8-04 & 5.59 & 2.40 & 1.32 & G020.6367$-$00.0452 & 0.3 & 3.6 & 0.055 & -0.250 & --- \\ 
18285844$-$1056069 & Mc8-05 & 5.57 & 2.08 & 1.23 & G020.6545$-$00.0562 & 0.9 & 2.2 & 0.179 & -0.329 & 108.6 \\ 
18290303$-$1053153 & Mc8-06 & 4.49 & 1.59 & 0.76 & G020.7053$-$00.0511 & 1.0 & 2.3 & -0.035 & -0.482 & --- \\ 
18385699$-$0606459 & St2-26 & 6.95 & 3.51 & 1.57 & G026.0708$-$00.0208 & 7.6 &  1.8  & 0.124 &  ---   & 92.3\\   
18390238$-$0605106 & St2-18 & 2.90 & 2.45 & 1.80 & G026.1044$-$00.0283 & 5.6 & 56.9  & 0.260 & $-$0.202 & 92.7\\ 
18390558$-$0604265 & St2-11 & 7.32 & 4.31 & 2.60 & G026.1215$-$00.0345 & 4.5 & 16.6  & 0.354 & $-$0.067 & 98.6\\  
18390776$-$0603203 & St2-08 & 5.23 & 2.26 & 1.08 & G026.1431$-$00.0343 & 3.4 &  1.5  & 0.049 &  ---   & --- \\  
18390805$-$0605244 & St2-14 & 4.82 & 2.48 & 1.23 & G026.1120$-$00.0510 & 4.7 &  4.9  & 0.159 & $-$0.219 & --- \\  
18391223$-$0553586 & St2-29 & 6.47 & 3.53 & 1.77 & G026.2891$+$00.0206 & 7.9 &  1.6  & 0.153 & 0.047  & --- \\  
18391470$-$0601366 & St2-05 & 5.24 & 2.00 & 0.97 & G026.1806$-$00.0465 & 1.4 &  1.5  & $-0$.051 &  ---  & --- \\  
18391489$-$0609272 & St2-28 & 5.60 & 3.07 & 1.61 & G026.0649$-$00.1076 & 7.8 &  3.4  & 0.121 & $-0$.289 & --- \\  
18391825$-$0602142 & St2-02 & 5.26 & 2.15 & 0.95 & G026.1782$-$00.0641 & 0.7 &  1.5  & $-0$.042 &  ---  & --- \\  
18391961$-$0600408 & St2-03 & 4.12 & 1.85 & 0.93 & G026.2038$-$00.0574 & 1.0 & 19.9  & 0.180 & $-0$.179 &102.7\\  
18391989$-$0601481 & St2-01 & 5.11 & 1.80 & 0.91 & G026.1886$-$00.0679 & 0.0 &  1.9  & $-0$.021 &  ---  & --- \\ 
18392161$-$0555197 & St2-21 & 8.86 & 1.99 & 2.76 & G026.2868$-$00.0240 & 6.3 &  2.7  & 0.158 & $-0$.152 & --- \\ 
18392461$-$0602138 & St2-04 & 4.50 & 1.82 & 0.96 & G026.1903$-$00.0876 & 1.1 &  6.6  & 0.088 & $-0$.401 & --- \\  
18392736$-$0607408 & St2-20 & 5.45 & 2.85 & 1.50 & G026.1150$-$00.1395 & 6.2 &  3.4  & 0.223 & $-0$.210 & --- \\  
18392891$-$0556435 & St2-17 & 5.82 & 2.07 & 1.25 & G026.2800$-$00.0617 & 5.4 &  5.5  & 0.053 & $-0$.209 & 50.2\\  
18392947$-$0557165 & St2-15 & 4.65 & 2.20 & 1.06 & G026.2732$-$00.0679 & 4.9 &  5.4  & 0.135 & $-0$.206 & --- \\  
18394340$-$0556290 & St2-27 & 5.88 & 1.95 & 0.90 & G026.3112$-$00.1130 & 7.7 &  3.1  & 0.716 &  ---   & --- \\  
18394635$-$0559473 & St2-23 & 4.51 & 1.71 & 0.83 & G026.2678$-$00.1492 & 6.7 &  5.9  & 0.115 & $-0$.126 & --- \\  
18534227$+$0041459 & Mc13-1 & 9.51 & 4.21 & 3.29 & G033.8104$-$00.1869 & 2.5 & 18.8 & 0.348 & 0.470 & --- \\ 
18534913$+$0038021 & Mc13-2 & 6.13 & 2.81 & 1.46 & G033.7671$-$00.2406 & 1.3 & --- & ---   & ---    & --- \\ 
18535060$+$0039015 & Mc13-3 & 6.01 & 2.68 & 1.27 & --- & ---  & --- & --- & --- & --- \\ 
18535240$+$0040172 & Mc13-4 & 5.89 & 3.67 & 1.95 & G033.8071$-$00.2351 & 1.5 & 3.6 & 0.103 & -0.216 & --- \\ 
18535249$+$0039313 & Mc13-5 & 2.67 & 3.03 & 1.62 & G033.7963$-$00.2417 & 1.6 & 144.9 & 0.184 & -0.257 & 72.3 \\ 
18540749$+$0039039 & Mc13-6 & 7.28 & 2.87 & 1.66 & G033.8179$-$00.3008 & 1.0 & 3.2 & 0.079 & -0.101 & 73.3 \\ 
\end{longtable}
\clearpage
\begin{longtable}{lcrrrrrrrrr}
  \caption{Observational results of the SiO maser search.} 
\hline  \hline
     &    & \multicolumn{4}{c}{SiO $J=1$--0 $v=1$}  & \multicolumn{4}{c}{SiO $J=1$--0 $v=2$} & \\
	 \cline{3-10} 
2MASS name  & No$^{*}$ &  $T_a$  &   $V_{\rm lsr}$  & L.F.  &  $rms$  &  $T_a$  &    $V_{\rm lsr}$ & L.F.  &  $rms$ & obs. date \\  
     &    & {\tiny (K)} & {\tiny (km s$^{-1}$)} & {\tiny (K km s$^{-1}$)} &  {\tiny (K)} & {\tiny (K)} & {\tiny (km s$^{-1}$)} & {\tiny (K km s$^{-1}$)} &  {\tiny (K)} & {\tiny (yymmdd.d)}\\
\hline
\endfirsthead
\hline \hline
     &    & \multicolumn{4}{c}{SiO $J=1$--0 $v=1$}  & \multicolumn{4}{c}{SiO $J=1$--0 $v=2$} & \\
	 \cline{3-10} 
2MASS name  & No$^{*}$ &  $T_a$  &   $V_{\rm lsr}$  & L.F.  &  $rms$  &      $T_a$  &    $V_{\rm lsr}$ & L.F.  &  $rms$ & obs. date \\  
     &    & {\tiny (K)} & {\tiny (km s$^{-1}$)} & {\tiny (K km s$^{-1}$)} &  {\tiny (K)} & {\tiny (K)} & {\tiny (km s$^{-1}$)} & {\tiny (K km s$^{-1}$)} &  {\tiny (K)} & {\tiny (yymmdd.d)}\\
\hline
\endhead
\hline
\endfoot
\hline
\multicolumn{11}{l}{$^{*}$ the star number in figure 1.} \\
\endlastfoot
18201544$-$1602142 & Mc4-05 &  --- &  --- &  --- & 0.049 & --- & --- & --- & 0.052 & 080427.2 \\
18202853$-$1602159 & Mc4-01 & --- & --- & --- & 0.038 & --- & --- & --- & 0.036 & 080428.2 \\
18203460$-$1606282 & Mc4-08 & --- & --- & --- & 0.051 & --- & --- & --- & 0.055 & 080427.2 \\
18203644$-$1602589 & Mc4-02 & 0.210 & 18.3 & 0.542 & 0.037 & --- & --- & --- & 0.035 & 080428.2 \\
18203850$-$1604163 & Mc4-03 & --- & --- & --- & 0.051 & --- & --- & --- & 0.056 & 080427.2 \\
18204070$-$1603412 & Mc4-04 & --- & --- & --- & 0.037 & --- & --- & --- & 0.035 & 080428.2 \\
18283530$-$1056364 & Mc8-01 & 0.932 & 74.6 & 3.285 & 0.051 & 0.673 & 73.7 & 2.601 & 0.053 & 080427.2 \\
18283912$-$1055526 & Mc8-02 & --- & --- & --- & 0.046 & --- & --- & --- & 0.048 & 080427.2 \\
18285207$-$1057576 & Mc8-03 & 0.312 & 101.9 & 0.826 & 0.052 & 0.311 & 102.9 & 0.927 & 0.050 & 080427.2 \\
18285402$-$1056453 & Mc8-04 & --- & --- & --- & 0.048 & --- & --- & --- & 0.050 & 080427.2 \\
18285844$-$1056069 & Mc8-05 & 0.221 & 108.6 & 0.374 & 0.050 & 0.273 & 108.6 & 0.846 & 0.054 & 080427.2 \\
18290303$-$1053153 & Mc8-06 & --- & --- & --- & 0.046 & --- & --- & --- & 0.050 & 080427.2 \\
18385699$-$0606459 & St2-26 & 0.258 &  94.2 &  0.589 & 0.065 &  0.330 & 90.4 & 0.582 & 0.068 & 060415.3 \\
18390238$-$0605106 & St2-18 & 0.367 &  92.7 &  4.199 & 0.062 &  ----  &  --- &  -----& 0.065 & 060412.3 \\
18390558$-$0604265 & St2-11 & 1.963 & 100.9 & 16.398 & 0.072 &  1.378 & 96.3 & 8.621 & 0.066 & 060412.3 \\
18390776$-$0603203 & St2-08 &  ---  &  ---  &    --- & 0.075 &  ---   &  --- &  ---  & 0.074 & 060413.3 \\
18390805$-$0605244 & St2-14 &  ---  &  ---  &    --- & 0.064 &  ---   &  --- &  ---  & 0.067 & 060412.3 \\
18391223$-$0553586 & St2-29 &  ---  &  ---  &    --- & 0.068 &  ---   &  --- &  ---  & 0.067 & 060415.3 \\
18391470$-$0601366 & St2-05 &  ---  &  ---  &    --- & 0.070 &  ---   &  --- &  ---  & 0.075 & 060413.3 \\
18391489$-$0609272 & St2-28 &  ---  &  ---  &    --- & 0.068 &  ---   &  --- &  ---  & 0.068 & 060415.3 \\
18391825$-$0602142 & St2-02 &  ---  &  ---  &    --- & 0.072 &  ---   &  --- &  ---  & 0.072 & 060413.3 \\
18391961$-$0600408 & St2-03 & 0.435 & 102.7 &  0.726 & 0.074 &  ---   &  --- &  ---  & 0.074 & 060413.3 \\
18391989$-$0601481 & St2-01 &  ---  &  ---  &    --- & 0.071 &  ---   &  --- &  ---  & 0.071 & 060413.3 \\
18392161$-$0555197 & St2-21 &  ---  &  ---  &    --- & 0.083 &  ---   &  --- &  ---  & 0.087 & 060416.3 \\
18392461$-$0602138 & St2-04 &  ---  &  ---  &    --- & 0.074 &  ---   &  --- &  ---  & 0.075 & 060413.3 \\
18392736$-$0607408 & St2-20 &  ---  &  ---  &    --- & 0.066 &  ---   &  --- &  ---  & 0.068 & 060415.3 \\
18392891$-$0556435 & St2-17 & 0.345 &  49.1 &  0.979 & 0.070 &  0.193 & 51.3 & 0.809 & 0.069 & 060415.3 \\
18392947$-$0557165 & St2-15 &  ---  &  ---  &    --- & 0.085 &  ---   &  --- &  ---  & 0.085 & 060416.3 \\
18394340$-$0556290 & St2-27 &  ---  &  ---  &    --- & 0.070 &  ---   &  --- &  ---  & 0.068 & 060415.3 \\
18394635$-$0559473 & St2-23 &  ---  &  ---  &    --- & 0.086 &  ---   &  --- &  ---  & 0.087 & 060416.3 \\
18534227$+$0041459 & Mc13-1 &  --- &  --- &  --- & 0.042 &  --- &  --- &  --- & 0.044 & 080427.1 \\
18534913$+$0038021 & Mc13-2 &  --- &  --- &  --- & 0.033 &  --- &  --- &  --- & 0.042 & 080427.2 \\
18535060$+$0039015 & Mc13-3 &  --- &  --- &  --- & 0.045 &  --- &  --- &  --- & 0.044 & 080427.1 \\
18535240$+$0040172 & Mc13-4 &  --- &  --- &  --- & 0.040 &  --- &  --- &  --- & 0.042 & 080427.2 \\
18535249$+$0039313 & Mc13-5 & 0.993 & 69.4 & 10.487 & 0.080 &  ---   &  --- &  ---  & 0.077 & 060413.4 \\ 
                   & Mc13-5 & 0.941 & 72.3 & 7.148 & 0.043 &  --- &  --- &  --- & 0.045 & 080427.1 \\
18540749$+$0039039 & Mc13-6 & 0.278 & 74.9 & 1.053 & 0.041 & 0.184 & 71.6 & 0.697 & 0.042 & 080427.2 \\
\end{longtable}

\clearpage
\begin{longtable}{lcrrrrrrrrr}
  \caption{Observational results of the H$_2$O maser search.} 
\hline  \hline
                 &    & \multicolumn{4}{c}{H$_2$O $6_{16}$--$5_{23}$}  & \multicolumn{4}{c}{Second Peak} &  \\
 \cline{3-10} 
2MASS name       & No$^{*}$ &  $Ta$  &   $V_{\rm lsr}$  & L.F.  &  $rms$  &      $Ta$  &    $V_{\rm lsr}$ & L.F.  &  $rms$ & obs. date \\  
                 &    & {\tiny (K)} & {\tiny (km s$^{-1}$)} & {\tiny (K km s$^{-1}$)} &  {\tiny (K)} & {\tiny (K)} & {\tiny (km s$^{-1}$)} & {\tiny (K km s$^{-1}$)} &  {\tiny (K)} & {\tiny (yymmdd.d)}\\
\hline
\endfirsthead
\hline \hline
                 &    & \multicolumn{4}{c}{H$_2$O $6_{16}$--$5_{23}$}  & \multicolumn{4}{c}{Second Peak} &  \\
 \cline{3-10} 
2MASS name       & No$^{*}$ &  $Ta$  &   $V_{\rm lsr}$  & L.F.  &  $rms$  &      $Ta$  &    $V_{\rm lsr}$ & L.F.  &  $rms$ & obs. date \\  
                 &    & {\tiny (K)} & {\tiny (km s$^{-1}$)} & {\tiny (K km s$^{-1}$)} &  {\tiny (K)} & {\tiny (K)} & {\tiny (km s$^{-1}$)} & {\tiny (K km s$^{-1}$)} &  {\tiny (K)} & {\tiny (yymmdd.d)}\\
\hline
\endhead
\hline
\endfoot
\hline
\multicolumn{11}{l}{$^{*}$ the star number shown in figure 1.}\\
\multicolumn{11}{l}{$^{\flat}$ contamination from Mc13-04.}\\
\multicolumn{11}{l}{$^{\dagger}$ contamination from Mc13-05.}\\
\endlastfoot
18201544$-$1602142 & Mc4-05 &  --- &  --- &  --- & 0.027 &  --- &  --- &  --- &  --- & 080430.1 \\
18202853$-$1602159 & Mc4-01 & 2.232 & 24.0 & 6.119 & 0.032 &  --- &  --- &  --- &  --- & 080430.1 \\
18203460$-$1606282 & Mc4-08 & 0.428 & -44.5 & 0.803 & 0.028 & 0.209 & 19.4 & 0.684 &  --- & 080430.1 \\
18203644$-$1602589 & Mc4-02 &  --- &  --- &  --- & 0.031 &  --- &  --- &  --- &  --- & 080430.1 \\
18203850$-$1604163 & Mc4-03 & 0.101 & 20.5 & 0.201 & 0.027 &  --- &  --- &  --- &  --- & 080430.1 \\
18204070$-$1603412 & Mc4-04 &  --- &  --- &  --- & 0.035 &  --- &  --- &  --- &  --- & 080430.1 \\
18283530$-$1056364 & Mc8-01 &  --- &  --- &  --- & 0.033 &  --- &  --- &  --- &  --- & 080430.2 \\
18283912$-$1055526 & Mc8-02 &  --- &  --- &  --- & 0.071 &  --- &  --- &  --- &  --- & 080430.2 \\
18285207$-$1057576 & Mc8-03 &  --- &  --- &  --- & 0.034 &  --- &  --- &  --- &  --- & 080430.2 \\
18285402$-$1056453 & Mc8-04 &  --- &  --- &  --- & 0.074 &  --- &  --- &  --- &  --- & 080430.2 \\
18285844$-$1056069 & Mc8-05 &  --- &  --- &  --- & 0.039 &  --- &  --- &  --- &  --- & 080430.2 \\
18290303$-$1053153 & Mc8-06 &  --- &  --- &  --- & 0.099 &  --- &  --- &  --- &  --- & 080430.2 \\
18385699$-$0606459 & St2-26 &  --- &  --- &  --- & 0.035 &  --- &  --- &  --- &  --- & 080430.2 \\
18390238$-$0605106 & St2-18 & 0.519 & 95.6 & 7.368 & 0.031 &  --- &  --- &  --- &  --- & 080430.1 \\
18390558$-$0604265 & St2-11 & 2.610 & 87.2 &28.605 & 0.032 &  --- &  --- &  --- &  --- & 080430.1 \\
18390776$-$0603203 & St2-08 & 0.167 & 86.7 & 1.583 & 0.031 &  --- &  --- &  --- &  --- & 080430.1 \\
18390805$-$0605244 & St2-14 &  --- &  --- &  --- & 0.032 &  --- &  --- &  --- &  --- & 080430.1 \\
18391223$-$0553586 & St2-29 &  --- &  --- &  --- & 0.033 &  --- &  --- &  --- &  --- & 080430.2 \\
18391470$-$0601366 & St2-05 &  --- &  --- &  --- & 0.031 &  --- &  --- &  --- &  --- & 080430.1 \\
18391489$-$0609272 & St2-28 &  --- &  --- &  --- & 0.033 &  --- &  --- &  --- &  --- & 080430.2 \\
18391825$-$0602142 & St2-02 &  --- &  --- &  --- & 0.030 &  --- &  --- &  --- &  --- & 080430.1 \\
18391961$-$0600408 & St2-03 & 0.112 & 106.8 & 0.336 & 0.034 &  --- &  --- &  --- &  --- & 080430.1 \\
18391989$-$0601481 & St2-01 &  --- &  --- &  --- & 0.032 &  --- &  --- &  --- &  --- & 080430.1 \\
18392161$-$0555197 & St2-21 &  --- &  --- &  --- & 0.033 &  --- &  --- &  --- &  --- & 080430.2 \\
18392461$-$0602138 & St2-04 &  --- &  --- &  --- & 0.035 &  --- &  --- &  --- &  --- & 080430.1 \\
18392736$-$0607408 & St2-20 &  --- &  --- &  --- & 0.042 &  --- &  --- &  --- &  --- & 080430.2 \\
18392891$-$0556435 & St2-17 &  --- &  --- &  --- & 0.033 &  --- &  --- &  --- &  --- & 080430.2 \\
18392947$-$0557165 & St2-15 &  --- &  --- &  --- & 0.034 &  --- &  --- &  --- &  --- & 080430.2 \\
18394340$-$0556290 & St2-27 &  --- &  --- &  --- & 0.037 &  --- &  --- &  --- &  --- & 080430.2 \\
18394635$-$0559473 & St2-23 &  --- &  --- &  --- & 0.035 &  --- &  --- &  --- &  --- & 080430.2 \\
18534227$+$0041459 & Mc13-1 & 2.476 & 45.4 & 5.225 & 0.032 &  --- &  --- &  --- &  --- & 080430.0 \\
18534913$+$0038021 & Mc13-2 &  --- &  --- &  --- & 0.032 &  --- &  --- &  --- &  --- & 080430.0 \\
18535060$+$0039015 & Mc13-3 & 0.717$^{\dagger}$ & 70.2$^{\dagger}$ & 2.163$^{\dagger}$ & 0.034 &  --- &  --- &  --- &  --- & 080430.0 \\
18535240$+$0040172 & Mc13-4 & 0.445 & 21.4 & 0.950 & 0.031 & 0.635$^{\dagger}$ & 70.2$^{\dagger}$ & 1.858$^{\dagger}$ &  --- & 080430.0 \\
18535249$+$0039313 & Mc13-5 & 1.495 & 70.2 & 5.011 & 0.036 & 0.197$^{\flat}$ & 20.4$^{\flat}$ & 0.404$^{\flat}$ &  --- & 080430.0 \\
18540749$+$0039039 & Mc13-6 &  --- &  --- &  --- & 0.033 &  --- &  --- &  --- &  --- & 080430.0 \\
\end{longtable}
\begin{longtable}{lrrll}
 \caption{Other names for observed objects.} 
\hline  \hline
2MASS name       & No$^{*}$ & D07$^{1}$ & MSX name & other names \& note \\ 
\hline
\endfirsthead
\hline \hline
2MASS name       & No$^{*}$ & D07$^{1}$ & MSX name & note \& other \& names  \\ 
\hline
\endhead
\hline
\endfoot
\hline
\multicolumn{5}{l}{$^{*}$ the star number in figure 1.} \\
\multicolumn{5}{l}{$^{1}$ \cite{dav07}.} \\
\multicolumn{5}{l}{$^{2}$ \cite{and02}. } \\
\endlastfoot
18201544$-$1602142 & Mc4-05 &   & G015.1542$-$00.5707 & \\ 
18202853$-$1602159 & Mc4-01 &   & ---                 & H$_2$O G15.18$-$0.62 \\
18203460$-$1606282 & Mc4-08 &   & G015.1288$-$00.6717 & M17WF 24$^2$ \\ 
18203644$-$1602589 & Mc4-02 &   & G015.1831$-$00.6482 & M17WF 26$^2$ \\ 
18203850$-$1604163 & Mc4-03 &   & ---                 & M17WF 29$^2$\\ 
18204070$-$1603412 & Mc4-04 &   & ---                 & \\
18283530$-$1056364 & Mc8-01 &   & G020.6034$+$00.0234 & IRAS 18258$-$1058 \\ 
18283912$-$1055526 & Mc8-02 &   & G020.6214$+$00.0151 & \\ 
18285207$-$1057576 & Mc8-03 &   & G020.6156$-$00.0476 & \\ 
18285402$-$1056453 & Mc8-04 &   & G020.6367$-$00.0452 & \\ 
18285844$-$1056069 & Mc8-05 &   & G020.6545$-$00.0562 & \\ 
18290303$-$1053153 & Mc8-06 &   & G020.7053$-$00.0511 & StRS 202 \\ 
18385699$-$0606459 & St2-26 &   & G026.0708$-$00.0208 & \\  
18390238$-$0605106 & St2-18 & 1 & G026.1044$-$00.0283 & IRAS 18363$-$0607 (17$''$) \\  
18390558$-$0604265 & St2-11 &49 & G026.1215$-$00.0345 & K4 \\  
18390776$-$0603203 & St2-08 & 9 & G026.1431$-$00.0343 & M5 IRAS 18364$-$0605  (26$''$) \\  
18390805$-$0605244 & St2-14 & 5 & G026.1120$-$00.0510 & M4 \\  
18391223$-$0553586 & St2-29 &   & G026.2891$+$00.0206 & \\  
18391470$-$0601366 & St2-05 &10 & G026.1806$-$00.0465 & M5 \\  
18391489$-$0609272 & St2-28 &   & G026.0649$-$00.1076 & \\  
18391825$-$0602142 & St2-02 &11 & G026.1782$-$00.0641 & M4 \\  
18391961$-$0600408 & St2-03 & 2 & G026.2038$-$00.0574 & M3 Cl Stephenson 2 2 IRC $-$10447  \\  
18391989$-$0601481 & St2-01 & 8 & G026.1886$-$00.0679 & K5 Cl Stephenson 2 4 \\  
18392161$-$0555197 & St2-21 &   & G026.2868$-$00.0240 & \\  
18392461$-$0602138 & St2-04 & 3 & G026.1903$-$00.0876 & M4 Cl Stephenson 2 10 \\  
18392736$-$0607408 & St2-20 &   & G026.1150$-$00.1395 & \\  
18392891$-$0556435 & St2-17 &22 & G026.2800$-$00.0617 & \\  
18392947$-$0557165 & St2-15 & 4 & G026.2732$-$00.0679 & \\  
18394340$-$0556290 & St2-27 &   & G026.3112$-$00.1130 & \\  
18394635$-$0559473 & St2-23 &   & G026.2678$-$00.1492 & IRAS 18370-0602  StRS 239  \\  
18534227$+$0041459 & Mc13-1 &   & G033.8104$-$00.1869 & IRAS 18511+0038 \\ 
18534913$+$0038021 & Mc13-2 &   & G033.7671$-$00.2406 & \\ 
18535060$+$0039015 & Mc13-3 &   & ---                 & \\               
18535240$+$0040172 & Mc13-4 &   & G033.8071$-$00.2351 & \\ 
18535249$+$0039313 & Mc13-5 &   & G033.7963$-$00.2417 & IRAS 18513+0035 NSV 11485\\ 
18540749$+$0039039 & Mc13-6 &   & G033.8179$-$00.3008 & \\ 
\end{longtable}
\end{document}